\documentclass[12pt,preprint]{aastex}

\def\simgt{\lower 2pt \hbox{$\, \buildrel {\scriptstyle >}\over{\scriptstyle \sim}\,$}}
\def\simlt{\lower 2pt \hbox{$\, \buildrel {\scriptstyle <}\over{\scriptstyle \sim}\,$}}

\begin{document}

\title{A {\it Chandra\/} Survey of the \hbox{X-ray} Properties of Broad
Absorption Line Radio-Loud Quasars}

\author{B.~P.~Miller,\footnotemark[1] ~~W.~N.~Brandt,\footnotemark[1]
  ~~R.~R.~Gibson,\footnotemark[2] ~~G.~P.~Garmire,\footnotemark[1]
  ~~and O.~Shemmer\footnotemark[3]}

\footnotetext[1]{Department of Astronomy and Astrophysics, The
  Pennsylvania State University, 525 Davey Laboratory, University
  Park, PA 16802; {\it bmiller, niel, garmire@astro.psu.edu\/}}

\footnotetext[2]{Department of Astronomy, University of Washington,
  Physics-Astronomy Bldg Room C319, Seattle, WA 98195; {\it
    rgibson@astro.washington.edu\/}}

\footnotetext[3]{Department of Physics, University of North Texas,
  1155 Union Circle, \#311427, Denton, TX 76203; {\it ohad@unt.edu\/}}

\begin{abstract}

This work presents the results of a {\it Chandra\/} study of 21 broad
absorption line (BAL) radio-loud quasars (RLQs). We conducted a {\it
  Chandra\/} snapshot survey of 12 bright BAL RLQs selected from
SDSS/FIRST data and possessing a wide range of radio and C~IV
absorption properties. Optical spectra were obtained nearly
contemporaneously with the Hobby-Eberly Telescope; no strong flux or
BAL variability was seen between epochs. In addition to the snapshot
targets, we include in our sample 9 additional BAL RLQs possessing
archival {\it Chandra\/} coverage. We compare the properties of
(predominantly high-ionization) BAL RLQs to those of non-BAL RLQs as
well as to BAL radio-quiet quasars (RQQs) and non-BAL RQQs for
context.

All 12 snapshot and 8/9 archival BAL RLQs are detected, with observed
\hbox{X-ray} luminosities less than those of non-BAL RLQs having
comparable optical/UV luminosities by typical factors of
4.1--8.5. (BAL RLQs are also \hbox{X-ray} weak by typical factors of
2.0--4.5 relative to non-BAL RLQs having both comparable optical/UV
and radio luminosities.)  However, BAL RLQs are not as \hbox{X-ray}
weak relative to non-BAL RLQs as are BAL RQQs relative to non-BAL
RQQs. While some BAL RLQs have harder X-ray spectra than typical
non-BAL RLQs, some have hardness ratios consistent with those of
non-BAL RLQs, and there does not appear to be a correlation between
\hbox{X-ray} weakness and spectral hardness, in contrast to the
situation for BAL RQQs. RLQs are expected to have \hbox{X-ray}
continuum contributions from both disk-corona and small-scale jet
emission. While the entire \hbox{X-ray} continuum in BAL RLQs cannot
be obscured to the same degree as in BAL RQQs, we calculate that the
jet is likely partially covered in many BAL RLQs. We comment briefly
on implications for geometries and source ages in BAL RLQs.

\end{abstract}


\section{Introduction}

Quasar outflows help regulate the accretion structure about the
central supermassive black hole and propagate kinetic energy into the
surrounding environment. Apparently the most extreme manifestation of
outflows observed in radio-quiet quasars (RQQs) is the blue-shifted
broad absorption lines (BALs) present in the rest-frame UV spectra of
$\simeq$18--26\% of RQQs (e.g., Hewett \& Foltz 2003). In an
orientation-based unification model, this fraction represents the
covering factor of the BAL wind that is common to RQQs. The high
polarization within BAL troughs (e.g., Ogle et al.~1999) supports
orientation models, while the general IR similarity of BAL and non-BAL
RQQs (e.g., Willott et al.~2003; Gallagher et al.~2007) argues against
competing ``dust-shroud'' evolutionary models. BAL RQQs are usually
weaker in X-rays than would be expected from their optical
luminosities (e.g., Gallagher et al.~2006; Gibson et al.~2009). The
X-ray spectra of BAL RQQs show clear evidence of \hbox{X-ray}
absorption, often complex, with intrinsic column densities
\hbox{$N_{\rm H} >$ 10$^{22}$ cm$^{-2}$} (e.g., Gallagher et
al.~2002). Although UV and X-ray absorption are clearly linked (e.g.,
Brandt, Laor, \& Wills 2000), the higher column density of the
\hbox{X-ray} absorber (e.g., Arav et al.~2003) suggests the
\hbox{X-ray} absorption arises interior to the UV BALs, perhaps in the
``shielding gas'' postulated by Murray et al.~(1995) and generated
naturally in the simulations of Proga et al.~(2000).

A lack of detected BALs in radio-loud\footnote{We follow the
  convention that ``radio-loudness'' ($R^{*}$) is defined by the ratio
  of monochromatic luminosities at rest-frame 5~GHz and
  2500~\AA~(e.g., Stocke et al.~1992), where optical/UV luminosities
  have been corrected for any strong intrinsic reddening. RQQs have
  $R^{*}<10$ while RLQs require at least $R^{*}>10$; we consider
  objects with $10<R^{*}<50$ to be radio-intermediate and those with
  $R^{*}>50$ to be definitively radio-loud.}  quasars (RLQs) led
to early suggestions that quasars could possess either a jet, or a BAL
wind, but not both simultaneously (e.g., Stocke et al.~1992). However,
an increasing number of individual BAL RLQs began to be discovered
(e.g., Becker et al.~1997; Brotherton et al.~1998; Wills et al.~1999;
Gregg et al.~2000; Ma~2002; Benn et al.~2005), and systematic optical
spectroscopic coverage of quasars detected in the {\it VLA\/} 1.4 GHz
FIRST survey (Becker et al.~1995) obtained by the FIRST Bright Quasar
Survey (FBQS; White et al.~2000) and the Sloan Digital Sky Survey
(SDSS; York et al.~2000) has increased the number of known radio-loud
BAL quasars to $\simgt$100 (e.g., Becker et al.~2000, 2001; Menou et
al.~2001; Shankar et al.~2008). The fraction of quasars with BALs does
decrease with increasing radio luminosity (e.g., Shankar et
al.~2008). Several of the discovered BAL RLQs have flat or convex
radio spectra and/or compact morphologies (e.g., Becker et al.~2000;
Liu et al.~2008; Montenegro-Montes et al.~2008), similar to the radio
properties of compact steep spectrum (CSS) or GHz peaked spectrum
(GPS) radio sources (e.g., O'Dea 1998). The association of BAL RLQs
with GPS/CSS radio sources, commonly presumed to be young (e.g.,
Stawarz et al.~2008 and references therein), has revived evolutionary
scenarios (e.g., Gregg et al.~2006), as has the prevalence of objects
with low-ionization BALs among dust-reddened quasars (Urrutia et
al.~2009) which are plausibly newly active (e.g., Urrutia et
al.~2008). Further, Zhou et al.~(2006) identify six BAL RLQs
($R^{*}\simlt250$ after correcting for intrinsic extinction) with high
brightness temperatures suggesting the nucleus is observed from a
polar perspective. The presence of BALs in low-inclination RLQs would
seem inconsistent with the quasi-equatorial disk-wind model often
applied to RQQs.

X-ray observations of BAL RLQs can provide insight into the nature of
the BAL outflow, through quantifying any \hbox{X-ray} weakness or
spectral hardening associated with BAL-linked
absorption. Unfortunately, there have been only a handful of
\hbox{X-ray} studies of BAL RLQs to date. Brotherton et al.~(2005)
conducted a {\it Chandra\/} study of five BAL RLQs and found that they
were \hbox{X-ray} weak but had relatively soft spectra, consistent
with complex absorption or a jet-dominated continuum. These sources
were of intermediate radio-loudness (with none having a dereddened
$R^{*}>100$), and three of the five were in the minority population of
low-ionization BAL quasars, which in RQQs display particularly strong
\hbox{X-ray} absorption (e.g., Green et al.~2001; Gibson et
al.~2009). Wang et al.~(2008) present {\it XMM-Newton} observations of
four BAL RLQs believed to be viewed at low inclinations (three of
which have low-ionization BALs and only one of which has $R^{*}>100$),
finding the two detected BAL RLQs to lack intrinsic \hbox{X-ray}
absorption and to possess normal \hbox{X-ray}
luminosities. \hbox{X-ray} studies of individual BAL RLQs include the
work of Schaefer et al.~(2006), who find J101614.25+520915.4 to be
X-ray weak with significant soft \hbox{X-ray} emission, and Miller et
al.~(2006), who detect variable \hbox{X-ray} absorption in the BAL RLQ
PG~1004+130 (but with an observed column density less than that of
most BAL RQQs) and suggest the power-law form and \hbox{X-ray}
weakness of the unabsorbed \hbox{X-ray} spectrum may indicate a
jet-dominated \hbox{X-ray} nuclear continuum. Statistical efforts to
understand the \hbox{X-ray} properties of BAL RLQs require a larger
sample covering a wide range of BAL and radio properties.

We here present results from a {\it Chandra\/} snapshot survey of a
well-defined sample of 12 BAL RLQs primarily selected from the SDSS
Data Release 3 (DR3) BAL quasar catalog of Trump et al.~(2006). The
objects were selected to be distinctly radio-loud ($R^{*}\simgt100$)
with strong C~IV absorption spanning a range of equivalent widths
(EW\footnote{We use positive values throughout for C~IV absorption EW;
  emission line properties are not considered in this work. All EW
  values are rest-frame.}) and velocities; both core-dominated and
lobe-dominated radio sources are included in the sample. For most of
these objects we were able to obtain optical spectra and photometry
with the Hobby-Eberly Telescope ({\it HET\/}) within $\sim$1--3
rest-frame weeks of the {\it Chandra\/} pointing, to check for BAL and
continuum variability. We also make use of {\it Chandra\/} archival
data for an additional 9 BAL RLQs (including those observed by SDSS in
DR4 and DR5), all of which have C~IV absorption EW~$>5$~\AA~and
$R^{*}\simgt50$. Even taking redshift censoring into account, our
sample is dominated by high-ionization BAL quasars, which represent
the majority of SDSS-selected BAL quasars (e.g., Trump et
al.~2006). We have carefully constructed comparison samples of non-BAL
RLQs, BAL RQQs, and non-BAL RQQs observed with SDSS/FIRST/{\it
  Chandra\/} in order to provide proper context for our results and to
enable interpretation of the physical nature of BAL outflows in
RLQs. Such comparisons are necessitated by the presence of a
radio-linked component in the \hbox{X-ray} emission of RLQs, apparent
both through increased X-ray luminosity (e.g., Worrall et al.~1987)
and X-ray spectral flattening (e.g., Wilkes \& Elvis et al.~1987) with
increasing radio loudness, and commonly presumed to arise in a
small-scale jet.

This paper is organized as follows: $\S$2 describes the selection of
the snapshot and archival BAL RLQ samples and the construction of
comparison samples, $\S$3 presents the {\it HET\/} optical and {\it
  Chandra\/} \hbox{X-ray} observations and provides notes on individual
objects, $\S$4 quantifies \hbox{X-ray} luminosities and spectral
properties relative to the comparison samples, $\S$5 discusses
physical interpretations of BAL RLQs, and $\S$6 summarizes the main
conclusions. A standard cosmology with \hbox{$H_{0}$ = 70 km
  s$^{-1}$~Mpc$^{-1}$}, ${\Omega}_{M}$~=~0.3, and
${\Omega}_{\Lambda}$~=~0.7 is assumed throughout. Unless otherwise
noted, errors are given as 90\% confidence intervals for one parameter
of interest (${\Delta}{\chi}^{2}$~=~2.71). Radio, optical/UV, and
X-ray monochromatic luminosities $l_{\rm r}$, $l_{\rm uv}$, and
$l_{\rm x}$ have units of log~ergs~s$^{-1}$~Hz$^{-1}$, at rest-frame
frequencies of 5~GHz, 2500~\AA, and 2~keV, respectively. Object names
are typically given as SDSS J2000.

\section{Sample properties}

Our sample of BAL RLQs consists of 21 objects, greatly increasing the
number of BAL RLQs with high-quality \hbox{X-ray} coverage. 12 of these BAL
RLQs have \hbox{X-ray} data from a {\it Chandra\/} snapshot survey (PI Garmire)
and 9 have archival {\it Chandra\/} coverage. 20/21 BAL RLQs are
detected in the 0.5--8~keV band. We also make use of comparison
samples of RLQs, BAL RQQs, and RQQs with {\it Chandra\/} coverage.

\subsection{Selection of snapshot BAL RLQs}

The targets for the {\it Chandra\/} snapshot survey were selected from
the Trump et al.~(2006) BAL quasar catalog, which includes SDSS
quasars with spectra as of DR3. To ensure consistent consideration of
BAL properties, only C~IV absorption measurements were used. The
Absorption Index ($AI$; Hall et al.~2002) was required to exceed
1500~km~s$^{-1}$ to remove borderline BALs from further
consideration. Note that $AI$ is defined from zero velocity with a
minimum velocity width of 1000~km~s$^{-1}$, and is consequently less
restrictive than the traditional Balnicity Index ($BI$; Weymann et
al.~1991); several objects in our sample have
$BI=0$~km~s$^{-1}$. Optical/UV luminosities were determined from SDSS
photometry (corrected for Galactic extinction) through redshifting the
composite quasar spectrum of Vanden Berk et al.~(2001), convolving it
with the SDSS filters, and then using the nearest magnitude (or
nearest two magnitudes) to $2500\times(1+z)$~\AA~to determine the
continuum flux (assuming an optical/UV power-law continuum slope of
${\alpha}_{\nu}=-0.5$, which is reasonable at these wavelengths). We
verified that alternative methods of calculating luminosities (e.g.,
the spectral-fitting method of Gibson et al.~2009) yield similar
results. 

These BAL quasars were then checked against the FIRST radio catalog to
generate a list of BAL RLQs. Because FIRST has angular resolution
sufficient to detect extended radio emission (when present) as
distinct sources in many cases, it is necessary to consider the nearby
environment to include all components (which could be some combination
of core, lobes, and jet) and determine the full radio flux. Candidate
matches were considered from all fields in which there was either a
FIRST source within $2''$ of the SDSS optical position, or two or more
FIRST sources within $90''$. All candidate fields were then examined
to exclude intruding background sources (often identifiable due to an
optical counterpart seen in the Digitized Sky Survey image). Radio
luminosities were calculated at rest-frame 5~GHz, assuming radio
power-law continuum slopes of ${\alpha}_{\nu}=-0.5$ for core
components and ${\alpha}_{\nu}=-1.0$ for extended components, when
present. Candidates for inclusion in the list of {\it Chandra\/}
snapshot targets were required to be distinctly radio-loud, defined as
having $R^{*}\simgt100$ and $l_{r}>33$. The optical spectra were
checked for obvious signs of intrinsic reddening (see $\S2.3$) to
ensure that the radio-loudness values were not significantly
artificially elevated.

The target list for the snapshot {\it Chandra\/} survey was then
constructed from the brightest (in SDSS $m_{\rm i}$) BAL RLQs. As can
be seen in Figure 1, the survey is substantially complete within DR3
BAL RLQs to $m_{\rm i}<18.6$. (The single DR3 object near $m_{\rm
i}=17.5$ lacking {\it Chandra} coverage is J144707.41+520340.0, which
was considered for inclusion in the target list but dropped as lowest
priority due to having the lowest absorption index,
$AI=1517$~km~s$^{-1}$.) One BAL RLQ (J112506.95$-$001647.6) with a
fainter $m_{\rm i}\simeq18.9$ was included based on showing extended
radio structure. One BAL RLQ (J102258.41+123429.7) with a post-DR3
SDSS spectrum (hence not listed in Trump et al.~2006) was selected
from the quasar catalog of Schneider et al.~(2007) based on showing
extended radio structure along with BAL absorption. The full snapshot
sample of BAL RLQs is listed in Table 1.

Gibson et al.~(2009) provide an SDSS BAL quasar catalog that covers
through DR5, and we make use of this to search for BAL RLQs with
archival {\it Chandra\/} coverage (see $\S2.2$) and to characterize
the BAL properties of the snapshot and archival samples (this catalog
was not available at the time of our {\it Chandra\/} target
selection). All but one (J074610.50+230710.8) of the snapshot BAL RLQs
are listed in Gibson et al.~(2009), and the listed BAL RLQs targeted
in the snapshot survey all have C~IV EW~$>5$~\AA. Since the spectral
fitting method and BAL definition in the catalog of Gibson et
al.~(2009) differ slightly from those used by Trump et al.~(2006),
minor inconsistencies in absorption properties and object inclusion
are to be expected.

\subsection{Selection of archival BAL RLQs}

Two lists of BAL RLQs were checked for archival {\it Chandra\/} ACIS
non-grating coverage; the first (134 BAL RLQs) was generated by
cross-matching quasars with C~IV absorption measurements from the BAL
catalog of Gibson et al.~(2009) with the FIRST catalog, in a manner
analogous to that outlined in $\S2.1$, while the second ($\sim$50 BAL
RLQs) was drawn from mentions of individual BAL RLQs in the
literature.\footnote{BAL RLQs identified in the following references
  were checked for archival {\it Chandra\/} coverage: Becker et
  al.~1997; Brotherton et al.~1998; Wills et al.~1999; Gregg et
  al.~2000; Becker et al.~2000; Becker et al.~2001; Menou et al.~2001;
  Brotherton et al.~2002; Lacy et al.~2002; Ma et al.~2002; Willott et
  al.~2002; Jiang \& Wang~2003; Benn et al.~2005; Brotherton et
  al.~2005; Gallagher et al.~2005; Urrutia et al.~2005; Brotherton et
  al.~2006; Gallagher et al.~2006; Gregg et al.~2006; Miller et
  al.~2006; Schaefer et al.~2006; Zhou et al.~2006; Just et al.~2007;
  Kunert-Bajraszewska et al.~2007; Liu et al.~2008; Montenegro-Montes
  et al.~2008.}  Naturally, BAL RLQs can appear in both of these
lists. Candidates were required to be definitively radio-loud
($R^{*}\simgt50$ and $l_{\rm r}>32$) with strong C~IV absorption
(EW~$>5$~\AA). These radio criteria are slightly less stringent than
those required of objects in the snapshot BAL RLQ sample, so as to
include potentially interesting BAL RLQs with existing \hbox{X-ray}
coverage, but still select objects comfortably on the radio-loud side
of the canonical $R^{*}=10$ border.

Other \hbox{X-ray} telescopes cannot match the angular resolution of
{\it Chandra\/}, important for minimizing background contamination
with faint sources, and many also have lower sensitivies and/or cover
a significantly different energy range. We searched the {\it
  XMM-Newton} archives for observations pointed to within 15$'$ of any
of the BAL RLQs described above, and find coverage of only three
objects that would meet our selection criteria: J081102.91+500724.5
(Wang et al.~2008), J101614.25+520915.4 (Schaefer et al.~2006), and
J151630.30$-$005625.5 (PI Boehringer). Adding archival BAL RLQs
observed with other \hbox{X-ray} telescopes would not notably increase
our sample size or alter our conclusions.

The archival BAL RLQ sample is listed in Table 2. The snapshot and
archival BAL RLQs together span a wide range of absorption and radio
properties, and constitute a reasonably representative sample of
definitively radio-loud BAL RLQs (Figure 2). As mentioned by previous
authors (e.g., Shankar 2008), it is rare for quasars to be
simultaneously strongly absorbed and strongly radio-loud, but our
sample includes a few such objects. The SDSS spectra of the BAL RLQs
(Figure 3) display a variety of BAL structures. The majority of the
sample BAL RLQs have compact morphologies at arc-second scales, but
4/21 show double-lobed structure and are dominated by extended radio
emission.

\subsection{Reddening}

Some BAL RLQs targeted by {\it Chandra\/} or mentioned in the
literature have unusual and extreme characteristics, and caution is
warranted before including such objects in a statistical study. In
particular, objects with heavy intrinsic reddening may have
artificially elevated apparent radio-loudness values. The BAL RLQs
J100424.88+122922.2 (Lacy et al.~2002; Urrutia et al.~2005) and
J155633.77+351757.3 (Becker et al.~1997; Brotherton et al.~2005) have
radio-loudness values below our selection criteria after correcting
for intrinsic reddening (J100424.88+122922.2 is also gravitationally
lensed), and are therefore excluded from the archival sample. Both
these objects have low-ionization BALs, as do two additional BAL RLQs
presented in Brotherton et al.~(2005) which are also strongly reddened
(such that their corrected radio loudness values are below our
selection threshold, although both were already excluded from
consideration here due to their low redshifts precluding observation
of their C~IV absorption properties); this is not unexpected, as
low-ionization BAL quasars are known to be generally redder than
high-ionization BAL quasars (e.g., Reichard et al.~2003).

We use the relative color indicator ${\Delta}(g-i)$ (calculated by
subtracting the median quasar color at a given redshift) to check for
large intrinsic reddening (e.g., Hall et al.~2006), keeping in mind
that RLQs generally show slightly redder colors than do RQQs (e.g.,
Ivezi{\'c} et al.~2002). The snapshot and archival BAL RLQs have
relative colors that are on average redder than those of SDSS quasars
(although most of our BAL RLQs do have relative colors within the
range spanned by 90\% of SDSS quasars) but consistent with those of
BAL RLQs in general\footnote{A Kolmogorov-Smirnov (KS) test comparing
  the BAL RLQs observed with {\it Chandra\/} to SDSS/FIRST BAL RLQs
  with $R^{*}>50$ and $EW>5\AA$ gives $p=0.29$ (comparing to BAL RLQs
  with $R^{*}>10$ and $EW>0\AA$ gives $p=0.12$), indicating that the
  distribution of colors for the snapshot and archival BAL RLQs is not
  significantly different from that of BAL RLQs in general. Comparing
  the snapshot and archival BAL RLQs to BAL RQQs gives $p=0.03$.}
(Figure 4). They do not appear to have strongly distorted
radio-loudness values.

The only established low-ionization BAL RLQ in our sample is the archival
object \\ J081426.45+364713.5, and although it does have a notably red
relative color its ${\Delta}(g-i)$ value is within the tail of the BAL
RLQ distribution and is significantly less than that of the strongly
reddened J155633.77+351757.3 (Figure 4). The CSS BAL RLQ
J104834.24+345724.9 suffers from intrinsic reddening (Willott et
al.~2002), but its corrected radio-loudness is still quite high, so it
is retained in our archival sample but with a dereddened optical
luminosity.

\subsection{Comparison samples}

In order to interpret the \hbox{X-ray} properties of BAL RLQs, it is
necessary also to analyze comparison samples of non-BAL RLQs (e.g., to
gauge the expected \hbox{X-ray} luminosities, including the
contribution from an unresolved jet to the \hbox{X-ray} nuclear
emission), of BAL RQQs (e.g., to provide context for \hbox{X-ray}
absorption relative to UV properties), and of non-BAL RQQs (e.g., to
give a baseline for measuring \hbox{X-ray} weakness in BAL RQQs). We
caution that the comparison samples we use are specifically chosen to
permit comparative investigation of our samples of BAL RLQs and should
not necessarily be used to infer general properties of non-BAL RLQs,
BAL RQQs, or non-BAL RQQs, particularly those having luminosities
outside of the ranges studied here. The optical/UV luminosities and
redshifts of the BAL RLQs observed with {\it Chandra\/} and of the
comparison samples are shown in Figure~5.

We constructed a comparison sample of RLQs by matching the SDSS DR5
Quasar Catalog (Schneider et al.~2007) to FIRST in a manner analogous
to that described in $\S$2.1, and then retaining RLQs with {\it
  Chandra\/} coverage with constraints of off-axis angle less than
12$'$, exposure greater than 1~ks, ACIS-S or ACIS-I used as the
detector, and no grating. This list was then filtered to include only
RLQs with $R^{*}>50$ and $l_{\rm r}>32$ so as to match the selection
criteria for the BAL RLQ archival sample. \hbox{X-ray} luminosities were
determined from {\it Chandra\/} count rates using the method described
in $\S3.2$. There are 68 RLQs selected in this manner, of which 67
(99\%) are detected in X-rays. Additional luminous RLQs were added
from the sample of Worrall et al.~(1987) based on {\it Einstein\/}
observations; after correcting to our chosen cosmology, we select
those RLQs with $l_{\rm uv}>31.3$, which yields a further 36 RLQs, 32
(89\%) with \hbox{X-ray} detections. The total RLQ comparison sample
comprises 104 RLQs, 99 (95\%) with \hbox{X-ray} detections. Although some of
the RLQs have redshifts too low to permit ready observation of the
C~IV region, the fraction of BAL RLQs is small enough (see references
in $\S$1) that any contamination of the comparison sample is minor and
does not impact later analysis; we often refer to the RLQ comparison
sample as ``non-BAL RLQs'' throughout.

A comparison sample of BAL RQQs is taken from the BAL catalog of
Gibson et al.~(2009), combining their Table 1 (absorption properties)
with their Table 5 (X-ray data). All high-ionization BAL RQQs with
{\it Chandra\/} coverage were selected, a total of 37 objects of which
28 (76\%) have \hbox{X-ray} detections. We also include those high-ionization
BAL RQQs lacking SDSS spectra (i.e., not available for inclusion in
the Gibson et al.~2009 catalog) from the Large Bright Quasar Survey
(LBQS; e.g., Foltz et al.~1987) observed with {\it Chandra\/} by
Gallagher et al.~(2006), an additional 15 objects, 13 (87\%) detected
by {\it Chandra\/}. The total BAL RQQ comparison sample comprises 52
BAL RQQs, 41 (79\%) with \hbox{X-ray} detections.

A comparison sample of non-BAL RQQs is taken from Gibson et
al.~(2008a); this sample has an excellent combination of size,
high-quality \hbox{X-ray} coverage, and well-characterized UV
properties. It is composed of the 139 non-BAL RQQs in their sample B,
which is made up of optically-selected quasars (SDSS objects targeted
exclusively based on FIRST or ROSAT properties excluded) with
serendipitous (off-axis angle constrained to be $1'<\theta<10'$) {\it
  Chandra\/} coverage having exposure $>2.5$~ks. These objects span a
redshift range of \hbox{$1.7<z<2.7$}, with the lower-limit set to
permit detection of C~IV absorption if present (and thereby exclude
BAL RQQs) and the upper limit set to permit direct measurement of the
2500~\AA~continuum flux. We also include 21 highly luminous non-BAL
RQQs from Just et al.~(2007), taking all objects in their ``clean''
sample with SDSS and {\it Chandra\/} data, to match better the
luminosity range of the BAL RLQs. The total RQQ comparison sample
comprises 160 RQQs, all of which have \hbox{X-ray} detections.

\section{Observations and Notes}

\subsection{{\it HET} observations}

We obtained optical photometry and spectroscopy of 10/12 of the
snapshot BAL RLQs near-contemporaneously with the {\it Chandra\/}
observations, using the queue-scheduled Hobby-Eberly Telescope (Ramsey
et al.~1998). The Low-Resolution Spectrograph (LRS; Hill~et~al.~1998)
was used for the spectroscopic observations, generally with a 1.5$''$
slit and the g2 grating, providing a resolution of $R\simeq867$
(sufficient for productive comparison to SDSS spectra, which have a
typical resolution of $R\simeq1800$). The HET data were reduced with
the Image Reduction and Analysis
Facility\footnote{http://iraf.noao.edu/iraf/web/} (IRAF) software
system using standard techniques, and the resulting spectra are
presented in Figure 6. (The {\it HET\/} spectrum for
J102258.41+123429.7 is not shown; unfortunately the BAL region fell
too close to the edge of the chip to provide a useful comparison to
the SDSS data.) None of the objects displays strong absorption-line
variability, although a few objects show minor changes in BAL
structure (see $\S3.3.1$); a more detailed discussion of BAL
variability in RLQs is deferred to Miller et al.~(in preparation). We
also obtained $R$-band images and looked for any flux variability via
comparison to field stars and galaxies; none of the observed objects
showed large ($>0.5$ magnitudes) variability. The {\it HET\/}
observing log is provided in Table 1.

\subsection{{\it Chandra} observations}

All snapshot BAL RLQ {\it Chandra\/} observations were carried out
using the Advanced CCD Imaging Spectrometer (ACIS; Garmire et
al.~2003) with exposure times of 4--7~ks. The targets were positioned
at the aimpoint of the S3 chip, and data were collected in Very Faint
mode. The pipeline processing includes automatic application of both
the ACIS charge-transfer inefficiency correction and the
time-dependent gain adjustment, and it is carried out using the
calibration database version CALDB v3.4.2. The data were analyzed
using CIAO version 4.0.2.

The archival BAL RLQ J081426.45+364713.5 has two {\it Chandra\/}
observations of comparable quality. These were stacked for the
purposes of determining source detection and extracting counts, and
the resulting increase in signal-to-noise is helpful for more
accurately determining the \hbox{X-ray} properties of this faint off-axis
source.

Source extraction for the BAL RLQ snapshot and archival objects and
for the non-BAL RLQ comparison sample (see $\S2.2$) {\it Chandra\/}
sources was performed using 90\% encircled-energy radii, using nearby
source-free regions for background determination. We evaluate source
detection through comparison of the observed aperture counts to the
95\% confidence upper limit for background alone. Where the number of
background counts is less than 10 (as applies in almost all cases) we
use the Bayesian formalism of Kraft et al.~(1991) to determine the
limit; else, we use equation 9 from Gehrels (1986). If the aperture
counts exceed the 95\% confidence upper limit we consider the source
detected and calculate the net counts by subtracting the background
from the aperture counts and then dividing by the encircled-energy
fraction; else, the source is considered undetected and the upper
limit is used. All snapshot BAL RLQs are detected, with net 0.5--8~keV
counts ranging from 17 to $\sim$170 (Table 1), and 8/9 archival BAL
RLQs are detected (Table 2). We confirmed source detections for the
BAL RLQs by running the CIAO {\it wavdetect\/} routine on
200$\times$200 square pixel images centered at the SDSS object
coordinates, with wavelet scales of 1, 1.41, 2, 2.83, 4, and 5.66
pixels; most sources are detected using a significance threshold of
$10^{-6}$, while J081426.45+364713.5 is detected (in the stacked image
only) using a significance threshold of $10^{-5}$.

All BAL RLQs were examined for variability within the {\it Chandra\/}
observation using the Gregory-Loredo algorithm implemented by
CIAO\footnote{http://cxc.harvard.edu/ciao/ahelp/glvary.html}. This
method filters by relevant good time intervals and accounts for any
dither near chip edges for off-axis sources. The probability that a
source is variable can be indicated with a variability index, ranging
from 0 to 10; most BAL RLQs had values of 0 (``definitely not
variable'') with only 3 objects having variability indices as high as
2 (``probably not variable''). Longer exposures could more tightly
constrain variability on ks timescales, while repeat observations
could assess variability between epochs.

\subsection{Notes on individual objects}

Optical/UV properties (including absorption characteristics) of the
BAL RLQs are listed in Table 3, while radio fluxes and spectral
indices are given in Table 4. Below, we briefly comment on interesting
aspects of the BAL RLQs.

\subsubsection{Snapshot BAL RLQs}

{\it J074610.50+230710.8\/} has a relatively large C~IV absorption
index of $AI=2955$~km~s$^{-1}$ (Trump et al.~2006), and has wide and
deep BAL-like absorption structure (Figures 3 and~6) despite being the
only snapshot BAL RLQ not included in the Gibson et al.~(2009) BAL
quasar catalog. The HET spectrum shows enhanced absorption in the
higher-velocity BAL (Figure 6), perhaps qualitatively consistent with
the tendency of BAL RQQs to vary within narrow discrete regions
(Gibson et al.~2008b). It is the reddest snapshot BAL RLQ with
$\Delta(g-i)\simeq1.1$ (the next reddest snapshot BAL RLQ has
$\Delta(g-i)\simeq0.6$). An archival {\it VLA\/} X-band image (program
AB862, observation date 1998-05-04) suggests the radio spectrum of
this compact-morphology source peaks near 5~GHz.

{\it J083749.59+364145.4\/} has particularly strong C~IV, Si~IV,
Ly~$\alpha$, and O~VI BALs. The C~IV EW of 34.6~\AA~is the largest in
the snapshot or archival BAL RLQ sample. It appears to have a
GHz-peaked (possibly variable) radio spectral shape and is unresolved
at milliarcsecond scales (Montenegro-Montes et al.~2008, 2009).

{\it J085641.58+424254.1\/} displays strong N~V absorption. The HET
spectrum suggests the C~IV emission line might be slightly variable.

{\it J092913.96+375742.9\/} (also FBQS J092913.9+375742) appears to be
resolved in an X-band VLA image (program AG0574, observation date
1999-07-12). There is a jet-like feature with a flux of 2.1 mJy
located $0.5''$ West of the core.

{\it J102258.41+123429.7\/} is resolved into a double-lobed morphology
by FIRST (see Figure~7), and the southern lobe shows extended diffuse
emission past the primary hotspot. The {\it Chandra\/} image does not
show any extended \hbox{X-ray} emission, but there are only $\sim20$ X-ray
source counts.

{\it J105416.51+512326.0\/} has a flat radio spectrum that steepens to
${\alpha}_{\rm r}=-0.35$ above 1.4~GHz.

{\it J112506.95$-$001647.6\/} is resolved into a double-lobed
morphology by FIRST (de Vries et al.~2006; see Figure~7). The {\it
Chandra\/} image does not show any extended \hbox{X-ray} emission; there are
$\sim80$ \hbox{X-ray} source counts. The primary C~IV absorption trough is at
low velocity and splits the emission line.

{\it J115944.82+011206.9\/} (also B1157+014) was identified as a BAL
RLQ by Menou~et~al.~(2001), who noted that in addition to the primary
low-velocity BAL there is an additional absorption trough near
8000~km~s$^{-1}$. The depth of this secondary absorption may have
increased slightly between the SDSS and HET observations. The radio
spectrum appears to be double-peaked (Montenegro-Montes et
al.~2008). The source shows symmetric jet-like extended emission on
milliarcsecond scales and a one-sided misaligned sequence of faint
knots stretching to $\sim100$ milliarcseconds (Montenegro-Montes et
al.~2009). J115944.82+011206.9 has the lowest $m_{\rm i}$ in the
snapshot sample and has sufficient \hbox{X-ray} counts ($\sim170$ from
\hbox{0.5--8~keV}) for basic spectral analysis (Figure 8). The
relatively hard \hbox{X-ray} spectrum suggests intrinsic absorption; a
neutral absorber has a best-fit column density of $N_{\rm H} =
3.2^{+2.9}_{-2.1}\times10^{22}$~cm$^{-2}$ with an unusual flat photon
index of $\Gamma=1.06^{+0.35}_{-0.33}$ required.

{\it J123411.73+615832.6\/} has an atypical BAL structure, with a deep
and wide trough that decreases gradually in depth until smoothly
meeting the base of the Si~IV emission line. Narrow redshifted C~IV
absorption is also present. A C-band VLA image (program AP450,
observation date 2003-02-27) indicates the radio spectral index is
${\alpha}_{\rm r}\simeq-0.5$.

{\it J133701.39$-$024630.3\/} has the highest measured \hbox{X-ray} hardness
ratio in the snapshot or archival BAL RLQ sample. A C-band VLA image
(program AG400, observation date 1994-01-08) suggests this is a
flat-spectrum RLQ with ${\alpha}_{\rm r}\simeq-0.1$.

{\it J141334.38+421201.7\/} (also FBQS J141334.4+421201) was
identified as a BAL RLQ by Becker et al.~(2000). The radio spectrum is
complex (Montenegro-Montes et al.~2008) while the morphology is
compact with a one-sided jet on milliarcsecond scales (Liu et
al.~2008).

{\it J162453.47+375806.6\/} is described in detail by Benn et
al.~(2005), and we use their value of $BI=2990$~km~s$^{-1}$ and
estimate $V_{\rm max}=28300$~km~s$^{-1}$ rather than taking
measurements from Gibson et al.~(2009) (for which BAL absorption was
integrated to 25000~km~s$^{-1}$). The large minimum ($V_{\rm
  min}=20560$~km~s$^{-1}$) and maximum velocities of the C~IV BAL in
this source are unusual for BAL RLQs and unique within our snapshot
and archival samples. There is also low-velocity absorption, described
by Benn et al.~(2005) as a mini-BAL (defined as total velocity range
$<2000$~km~s$^{-1}$; the mini-BAL is shaded gray along with the
primary BAL in Figure 6 for identification). The radio spectrum is
GHz-peaked (steep at high frequencies) and milliarcsecond imaging
reveals a one-sided jet (Benn et al.~2005; Montenegro-Montes et
al.~2008, 2009).

\subsubsection{Archival BAL RLQs}

{\it J020022.01$-$084512.0\/} (also FBQS J0200$-$0845) was identified
as a BAL RLQ by Becker et al.~(2001). It was observed serendipitously
in an $\sim$18~ks ACIS-I image (ObsID 3265; PI Ebeling) and is
discussed by Gallagher et al.~(2005). The source has a radio-loudness
value of $R^{*}=48$, on the border for inclusion in the archival
sample.

{\it FBQS J0256$-$0119\/} was identified as a BAL RLQ by Becker et
al.~(2001). Flux measurements by Montenegro-Montes et al.~(2008)
indicate a steep radio spectrum; those authors also note the increased
flux measured by FIRST relative to NVSS may be due to variability. The
$\sim$5~ks ACIS-S observation shows FBQS J0256$-$0119 to be \hbox{X-ray} weak
but with a soft spectrum (Brotherton et al.~2005). We do not have
access to photometric magnitudes for this object, so we take the lack
of intrinsic reddening noted by Brotherton et al.~(2005) as
justification to set the relative color ${\Delta}(g-i)=0$.

{\it J081426.45+364713.5\/} has a low radio luminosity ($l_{\rm
r}=32.7$) and the reddest relative color [${\Delta}(g-i)=1.1$] in the
snapshot or archival BAL RLQ sample. The optical spectrum shows deep
and wide BALs in both high and low-ionization lines (Trump et al.~2006
categorize it as an FeLoBAL) and only weak emission lines. It was
observed serendipitously in two $\sim$10~ks ACIS-I exposures (ObsID
3436, 3437; PI Fox) and is \hbox{X-ray} weak. 

{\it J091951.29+005854.9\/} has a non-zero $BI=673.1$~km~s$^{-1}$ and
a C~IV EW of 6.9~\AA, but the BAL is relatively narrow and the
absorption index is low ($AI=1268$~km~s$^{-1}$). The radio loudness is
also borderline for our sample ($R^{*}=51$). It was observed
serendipitously in a $\sim$5~ks ACIS-S image (ObsID 7056; PI Murray)
but is not detected.

{\it J100726.10+124856.2\/} (also PG 1004+130) is a low-redshift
($z=0.24$) RLQ in which BALs were discovered by Wills et al.~(1999);
we use their values of $BI=850$~km~s$^{-1}$ and $V_{\rm
  max}=10000$~km~s$^{-1}$ since the SDSS spectrum does not cover the
C~IV region. It is also a hybrid-morphology radio source
(Gopal-Krishna \& Wiita 2000), with an edge-brightened lobe opposite a
broadening edge-darkened jet. It is perhaps the best-studied BAL RLQ
at \hbox{X-ray} frequencies: deep {\it XMM-Newton\/} and {\it Chandra\/}
observations show \hbox{X-ray} absorption variability and also reveal X-ray
jet emission (Miller et al.~2006). PG~1004+130 is \hbox{X-ray} weak relative
to comparable non-BAL RLQs.

{\it J104834.24+345724.9\/} (also 4C +35.23) is a CSS RLQ with a C~IV
BAL identified by Willott et al.~(2002). It is radio luminous, and
even after correcting for some intrinsic reddening it remains notably
radio-loud (Kunert-Bajraszewska et al.~2007). It was targeted by {\it
Chandra\/} in a $\sim$5~ks ACIS-S observation (ObsID 9320; PI
Kunert-Bajraszewska) and is detected with a hard \hbox{X-ray} spectral shape.

{\it J122033.87+334312.0\/} (also 3C~270.1) is a double-lobed
steep-spectrum RLQ with the second-highest radio-loudness
($\log{R^{*}}=4.2$) in the snapshot or archival BAL RLQ
sample. Low-velocity C~IV absorption has been known to be present in
this object for some time (e.g., Anderson et al.~1987) although it has
not necessarily been described as a BAL quasar; however, the balnicity
index measured from the SDSS data is non-zero ($BI=52.5$; Gibson et
al.~2009). J122033.87+334312.0 was observed serendipitously in a
$\sim$3~ks ACIS-S exposure (ObsID 2118; PI Cagnoni).

{\it J131213.57+231958.6\/} (also FBQS J131213.5+231958) was
identified as a BAL RLQ by Becker et al.~(2000) and shows a wide and
deep C~IV absorption trough that extends to 25000~km~s$^{-1}$ (from
FBQS data; the DR7 SDSS spectrum does not have sufficient short
wavelength coverage to see the BAL). It shows two-sided extended radio
emission on milliarcsecond scales but is core dominated (Jiang et
al.~2003) and likely variable (Montenegro-Montes et al.~2008); these
radio characteristics do not provide a self-consistent orientation
measure. Liu et al.~(2008) suggest this object is similar in some
respects to CSS sources. The $\sim$5~ks ACIS-S observation (Brotherton
et al.~2005) shows it to be \hbox{X-ray} weak but with an \hbox{X-ray} spectrum
actually somewhat softer than is typical of non-BAL RLQs.

{\it LBQS 2211$-$1915\/} was identified as a ``marginal'' BAL quasar
by Weymann et al.~(1991), with a non-zero but low $BI=27$~km~s$^{-1}$,
and is radio-loud based on an NVSS flux measurement. The $\sim$6~ks
ACIS-S observation (Gallagher et al.~2006) shows it to be \hbox{X-ray} weak
relative to non-BAL RLQs. We estimate the relative color to be
${\Delta}(g-i)=0.32$ based on data from Gallagher et al.~(2007).

\section{Data Analysis}

Because RLQs are generally more \hbox{X-ray} luminous than comparable RQQs
(e.g., Worrall et al.~1987), a direct comparison of the X-ray
properties of BAL RLQs to those of BAL RQQs is of limited value. To
gain additional insight, we quantify the degree to which BAL RLQs are
X-ray weak relative to non-BAL RLQs, then compare this to the X-ray
decrement for BAL RQQs relative to non-BAL RQQs. Much of the X-ray
weakness of BAL RQQs may be explained by low-energy \hbox{X-ray} absorption
(e.g., Gallagher et al.~2006), which can produce \hbox{X-ray} spectra that
are harder than typical; examination of the \hbox{X-ray} spectral properties
of BAL RLQs compared with those of non-BAL RLQs can clarify whether a
similar effect typically applies to BAL RLQs.

\subsection{Calculation of \hbox{X-ray} hardness ratios and luminosities}

When insufficient counts are available for productive spectral
modeling, as is unfortunately the case for the majority of our data,
the relative contributions of hard and soft \hbox{X-ray} emission to the
overall spectrum can be assessed from the hardness ratio \hbox{$HR =
  (H-S)/(H+S)$}, where $H$ and $S$ are the net hard-band (2--8~keV)
and soft-band (0.5--2~keV) counts, respectively. Large values of the
hardness ratio can indicate intrinsic absorption, or alternatively an
unusually flat power-law (or both effects together). Since all our
data are taken from {\it Chandra\/} (including the non-BAL RLQ, BAL
RQQ, and non-BAL RQQ comparison samples) we can compare hardness
ratios without concern for intrumental cross-calibration
effects. Hardness ratios for objects observed with one of the
front-illuminated CCDs have been adjusted by subtracting 0.14 to
enable direct comparison to the hardness ratios for the
back-illuminated CCDs (such as S3, which covers the ACIS-S aimpoint).

The probability distribution for the hardness ratio can be calculated
using the Bayesian formalism detailed in Jin et al.~(2006), using a
uniform prior (i.e., their equation 13). The maximum-likelihood
hardness ratio is simply $(H-S)/(H+S)$. We use this method to
calculate 1$\sigma$ errors on the value of $HR$, defined such that
68\% of the area above (below) the maximum likelihood hardness ratio
is enclosed within the range of the upper (lower) bound (e.g., Wu et
al.~2007). Where the total number of counts exceeds 100 symmetric
errors are calculated from equation 8 of Jin et al.~(2006).

X-ray luminosities are calculated from the 0.5--8~keV count rates,
which are converted to observed-frame 2~keV flux densities with
PIMMS\footnote{http://cxc.harvard.edu/toolkit/pimms.jsp}, in all cases
assuming Galactic absorption and a power-law spectrum with
$\Gamma$=1.5. This model is typical of RLQs: for example, Reeves \&
Turner (2000) found $\langle{\Gamma}\rangle$~=~1.66 with
$\sigma$~=~0.22 for an {\it ASCA\/} sample of 35 RLQs, while Page et
al.~(2005) determined $\langle{\Gamma}\rangle$~=~1.55 with
$\sigma$~=~0.29 for an {\it XMM-Newton\/} sample of 16 RLQs at
$z>~$2. However, reasonable alternate choices for $\Gamma$ have only a
few percent impact upon the calculated \hbox{X-ray} fluxes. Count rates for
archival observations were converted to flux densities using the
calibration appropriate to that cycle, in order to account for the
temporal changes in ACIS sensitivity. The ACIS-I model in PIMMS was
used for all front-illuminated chips.

The bandpass-corrected \hbox{X-ray} luminosities $l_{\rm x}$ are given in
units of log~ergs~s$^{-1}$~Hz$^{-1}$ at rest-frame 2~keV in Table
5. We also calculate \hbox{X-ray} luminosities $l_{\rm x,S}$ and $l_{\rm
x,H}$ at rest-frame 2~keV determined from the soft and hard-band count
rates, respectively, in order to investigate the influence of spectral
shape (and intrinsic absorption) on the \hbox{X-ray} luminosity.

\subsection{Relative \hbox{X-ray} luminosities}

X-ray and optical/UV luminosities are correlated in quasars, and so to
evaluate the degree of \hbox{X-ray} weakness in BAL quasars it is
necessary to compare to non-BAL quasars of similar optical/UV
luminosity. Extensive studies (e.g., Avni \& Tananbaum 1986; Strateva
et al.~2005; Steffen et al.~2006; Just et al.~2007; Kelly et al.~2007)
have demonstrated that \hbox{X-ray} luminosity in non-BAL RQQs may be
parameterized as $l_{\rm x}\propto\beta{\times}l_{\rm uv}$, where
$\beta \simeq 0.6-0.8$ (i.e., the linear ratio of monochromatic
optical/UV luminosity to monochromatic \hbox{X-ray} luminosity
increases with increasing $l_{\rm uv}$). There is continuing debate
(e.g., Just et al.~2007; Kelly et al.~2007) as to whether the
optical/UV-to-X-ray properties of individual RQQs are also
significantly dependent upon redshift, but for our purposes the
$l_{\rm x}(l_{\rm uv})$ parameterization is fully satisfactory to
explore the large deviations from predicted \hbox{X-ray} luminosity
that are seen in BAL quasars. We make use of the relation $l_{\rm x} =
0.636{\times}l_{\rm uv}+7.055$ (a linear fit to log luminosities)
found by Just et al.~(2007) taking $l_{\rm x}$ as the dependent
variable and fitting their large sample of non-BAL RQQs using the
Astronomy Survival Analysis Package (ASURV; Lavalley et
al.~1992). Using the Bayesian maximum-likelihood method of Kelly
(2007), which accounts for both upper limits and errors (we presume
uncertainties are dominated by typical quasar variability; cf. $\S3.5$
of Gibson et al.~2008), and fitting our comparison sample of RQQs
yields a similar relation\footnote{Here and for subsequent model fits
  the quoted parameter values are the median of draws from the
  posterior distribution and the errors are 1$\sigma$.} of
\hbox{$l_{\rm x} = (0.574\pm0.057){\times}l_{\rm
    uv}+(8.995\pm1.772)$}.

As an initial step toward understanding the \hbox{X-ray} luminosities of BAL
RLQs, we fit $l_{\rm x}(l_{\rm uv})$ for non-BAL RLQs as for non-BAL
RQQs, finding a best-fit correlation of \\ \hbox{$l_{\rm x} =
  (0.905\pm0.079){\times}l_{\rm uv}-(0.813\pm2.459)$}, with
significant scatter (Figure 9a). The majority of snapshot and archival
BAL RLQs have \hbox{X-ray} luminosities less than those of non-BAL RLQs with
comparable optical/UV luminosities, typically by a factor of 4.1--8.5
(median 6.6). However, the difference between observed and predicted
X-ray luminosity in BAL RLQs is not as extreme as that for BAL RQQs
relative to non-BAL RQQs (Figure 9b); here the difference is typically
a factor of 2.8--34.0 (median 11.4) but exceeds 40 for $\simeq$15\% of
BAL RQQs. The sample BAL RLQs also tend to be \hbox{X-ray} weaker than
non-BAL RQQs at low optical/UV luminosities.

The \hbox{X-ray} luminosities of non-BAL RLQs can also be parameterized as a
function of radio luminosity, with a best-fit result for the
comparison non-BAL RLQ sample of \\ \hbox{$l_{\rm x} =
  (0.617\pm0.043){\times}l_{\rm r}+(6.328\pm1.480)$} (Figure 9c). Some
BAL RLQs again tend to fall below the non-BAL RLQ correlation,
although to a lesser degree, while some are matched in radio and X-ray
properties to comparable non-BAL RLQs. Possibly the reduced offset in
$l_{\rm x}(l_{\rm r})$ for BAL RLQs reflects not only \hbox{X-ray} weakness
but also lower radio-loudness values for BAL RLQs than for the
comparison non-BAL RLQs; the median value of $\log{R^{*}}$ is 2.2 for
the BAL RLQs and 3.0 for the RLQs. The outlier with high radio
luminosity and low \hbox{X-ray} luminosity is the CSS source
J104834.24+345724.9 (see also $\S$5.2). Fitting \hbox{X-ray} luminosity as a
joint function of optical/UV {\it and\/} radio luminosity yields a
relation with reduced scatter: \hbox{$l_{\rm x} =
  (0.472\pm0.085){\times}l_{\rm uv}+(0.413\pm0.054){\times}l_{\rm
    r}-(1.392\pm1.192)$} (Figure 9d). Due to the similar coefficients
BAL RLQs have essentially averaged offsets from $l_{\rm x}(l_{\rm
  uv})$ and $l_{\rm x}(l_{\rm r})$ for $l_{\rm x}(l_{\rm uv}, l_{\rm
  r})$. Obviously more sophisticated models are possible, but this
provides a useful quantitative measure of \hbox{X-ray} luminosity in BAL RLQs
relative to RLQs taking into account both optical/UV and radio
properties.

The difference between the observed \hbox{X-ray} luminosity in BAL RLQs and
that predicted from non-BAL RLQs with comparable optical/UV
luminosities, \hbox{${\Delta}l_{\rm x,uv} = l_{\rm x} - l_{\rm
    x}(l_{\rm uv})$}, is plotted as a histogram in Figure 10a. The
scatter for non-BAL RLQs (Figure 10b) is smaller than the degree to
which BAL RLQs are \hbox{X-ray} weak. However, BAL RLQs do not extend to
extreme values of \hbox{X-ray} weakness (log offsets of $<-1$), as do some
BAL RQQs (Figure 10c; here ${\Delta}l_{\rm x,uv}$ is calculated from
the Just et al.~2007 relation for non-BAL RQQs) relative to non-BAL
RQQs (Figure 10d). Note that the underlying distribution of
${\Delta}l_{\rm x,uv}$ for BAL RQQs is even \hbox{X-ray} weaker than the
histogram in Figure 10c suggests, as there are a large number of X-ray
upper limits. There appears to be a limit to how \hbox{X-ray} weak BAL RLQs
can become. The Kaplan-Meier estimates of the median ${\Delta}l_{\rm
  x,uv}$ values are $-0.82$, $-0.03$, and $-1.06$ for BAL RLQs,
non-BAL RLQs, and BAL RQQs, respectively. A Peto-Prentice two-sample
test indicates that the distribution of ${\Delta}l_{\rm x,uv}$ values
for BAL RLQs is significantly different from that of non-BAL RLQs
(test statistic 6.621, $p<5\times10^{-5}$) and BAL RQQs (test
statistic 2.249, $p=0.02$). Figure 10e shows a histogram of the
difference between the observed \hbox{X-ray} luminosity in BAL RLQs and that
predicted from non-BAL RLQs with comparable optical/UV {\it and\/}
radio luminosities, \hbox{${\Delta}l_{\rm x} = l_{\rm x} - l_{\rm
    x}(l_{\rm uv}, l_{\rm r})$}. BAL RLQs are a factor of 2.0--4.5
(median 3.2) weaker in X-rays than comparable non-BAL RLQs. The
results are similar if the relative \hbox{X-ray} luminosity is instead
calculated from the soft or hard-band luminosities, $l_{\rm x,S}$ and
$l_{\rm x,H}$ (Figures 10f and 10g).

There is a general trend in quasars relating C~IV absorption to X-ray
weakness (e.g., Brandt, Laor, \& Wills 2000; Laor \& Brandt 2002;
Gallagher et al.~2006). We plot the relative \hbox{X-ray} luminosity
for BAL RLQs [presented as ${\Delta}{\alpha}_{\rm ox}$ for ease of
  comparison to previous work, where \hbox{${\alpha}_{\rm ox} =
    0.384{\times}(l_{\rm x}-l_{\rm uv})$} and
  \hbox{${\Delta}{\alpha}_{\rm ox} = {\alpha}_{\rm ox} - {\alpha}_{\rm
      ox}(l_{\rm uv})$} with ${\alpha}_{\rm ox}(l_{\rm uv})$
  calculated from the $l_{\rm x}(l_{\rm uv})$ relations given above]
versus C~IV EW (Figure 11a) and maximum outflow velocity (Figure
11b). It is apparent that even BAL RLQs with large C~IV equivalent
widths (10--40~\AA) do not have ${\Delta}{\alpha}_{\rm ox}<-0.5$, as
do many strongly absorbed BAL RQQs. BAL RLQs appear to follow the
correlation between maximum outflow velocity and relative X-ray
luminosity that holds for BAL RQQs, but only to a limiting value of
${\Delta}{\alpha}_{\rm ox}$, near which BAL RLQs are observed with a
wide range of outflow velocities.

Additional context for interpreting the relative X-ray luminosities of
some BAL RLQs is provided by their radio morphologies or spectral
properties, which can constrain source inclination (e.g., Wills \&
Brotherton 1995) or age (e.g., Stawarz et al.~2008). Lobe-dominated
BAL RLQs (nested symbols in Figures 9 and 11), which presumably lie at
larger angles to the line of sight than do core-dominated BAL RLQs,
show a range of behavior: J112506.95$-$001647.6 and
J122033.87+334312.0 are actually X-ray bright relative to non-BAL RLQs
with similar optical/UV luminosities, whereas J102258.41+123429.7 and
particularly J100726.10+124856.2 (PG 1004+130) are \hbox{X-ray}
weak. J122033.87+334312.0 becomes \hbox{X-ray} weak when radio
luminosity is also taken into account, and perhaps the \hbox{X-ray}
brightness of J112506.95$-$001647.6 reflects its relatively weak C~IV
BAL (low absorption index and maximum velocity). Plausibly young
objects [including the GPS sources J083749.59+364145.4 and
  J162453.47+375806.6, and the CSS source J104834.24+345724.9
  (4C+35.23)] seem to have strong BALs and to be X-ray weak, but
additional data are required to investigate such trends in detail.

\subsection{X-ray spectral characteristics}

Most of the quasars in the snapshot and archival BAL RLQ samples and
in the non-BAL RLQ, BAL RQQ, and non-BAL RQQ comparison samples lack
sufficient counts for productive spectral fitting, so we investigate
basic \hbox{X-ray} spectral properties using hardness ratios (see
$\S$4.1). We are primarily interested in whether BAL RLQs show
evidence for intrinsic \hbox{X-ray} absorption. Absorption by a neutral
column will preferentially remove soft \hbox{X-ray} emission and lead to
greater values of $HR$, although this effect can be diluted by complex
(partial covering or ionized) absorption, such as is established to
occur in BAL RQQs (e.g., Gallagher et al.~2002, 2006) and has been
suggested for BAL RLQs (e.g., Brotherton et al.~2005). Absorption
spectral effects are also diluted by increasing redshift pushing the
rest-frame soft band to lower observed-frame energies.

The non-BAL RLQs in our comparison sample have relatively uniform
hardness ratios that do not appear strongly dependent upon redshift
(Figure 12a), suggesting that the spectra for these RLQs are generally
dominated by a simple power-law component with a standard photon index
and insignificant intrinsic absorption. Many BAL RLQs have hardness
ratios similar to those of non-BAL RLQs, but there are several BAL
RLQs with harder \hbox{X-ray} spectra, although none with measured
$HR>0.2$. A Peto-Prentice two-sample test indicates that the
distribution of $HR$ values for BAL RLQs is significantly different
from that of non-BAL RLQs (test statistic 3.704,
$p=2\times10^{-4}$). BAL RQQs typically have harder \hbox{X-ray} spectra than
non-BAL RQQs (the distributions are statistically different, with
$p<5\times10^{-5}$) and can have extreme hardness ratios (Figure
12b). The distribution of hardness ratios for the snapshot and
archival sample of BAL RLQs is not statistically inconsistent with
that of BAL RQQs (test statistic 1.396, $p=0.16$). The Kaplan-Meier
estimates of the median/mean $HR$ values for the BAL RLQs, non-BAL
RLQs, BAL RQQs, and non-BAL RQQs in our samples are $-0.40$, $-0.52$,
$-0.26$, and $-0.57$, respectively; the distribution for BAL RLQs is
skewed, with a Kaplan-Meier estimate of the mean $HR$ of $-0.34$. The
slightly higher median $HR$ for non-BAL RLQs relative to non-BAL RQQs
might be expected from prior \hbox{X-ray} spectral studies, but the
distributions for our comparison samples of non-BAL RLQs and non-BAL
RQQs are not statistically inconsistent ($p=0.23$; possibly the more
radio-luminous RLQs observed by {\it Einstein\/} would have slightly
larger {\it Chandra\/} hardness ratios than the RLQs plotted
here). For reference, a photon index of $\Gamma=2$ approximately
corresponds to $HR=-0.6$ and $\Gamma=1.7$ to $HR=-0.5$.

Five BAL RLQs have particularly hard \hbox{X-ray} spectral shapes
(with $HR>-0.2$) relative to both RLQs and other BAL RLQs; these
objects include J092913.96+375742.9 and J115944.82+011206.9 (both of
which apparently have small-scale radio jet emission),
J133701.39$-$024630.3, J100726.10+124856.2 (PG 1004+130, for which the
{\it XMM-Newton} spectrum is softer), and J104834.24+345724.9 (the CSS
source 4C+35.23). Although this study is not designed to investigate
the X-ray spectral properties of various subcategories of BAL RLQs, we
note briefly that the GPS sources J083749.59+364145.4 and
J162453.47+375806.6 have soft \hbox{X-ray} spectral shapes, and the
FeLoBAL J081426.45+364713.5 has an intermediate $HR=-0.29$, slightly
harder than the median for the BAL RLQs studied here. 

The correlation between \hbox{X-ray} weakness and hardness ratio in BAL RQQs
(Figure 12d) is reflective of (often complex) absorption reducing the
soft-band \hbox{X-ray} flux in BAL RQQs (e.g., Gallagher et al.~2006). A
similar trend is not obviously apparent for BAL RLQs (Figure 12c), for
which there are several \hbox{X-ray} weak objects with low hardness ratios
(or soft \hbox{X-ray} spectra), and essentially no BAL RLQs with
${\Delta}l_{\rm x}(l_{\rm uv})\simlt-1$. Note that our observations
are sensitive to low values of ${\Delta}l_{\rm x}(l_{\rm uv})$ (only
one undetected BAL RLQ is not plotted, and many of the rest could be
detected if they were even \hbox{X-ray} weaker by a linear factor of 5--10;
see net counts in Tables 1 and 2); the sample simply lacks notably
X-ray weak BAL RLQs. It does not appear possible to ascribe X-ray
weakness in BAL RLQs to intrinsic absorption (with properties as in
BAL RQQs) obscuring the {\it entire\/} nuclear \hbox{X-ray} continuum source,
although such an interpretation could hold for some particular BAL
RLQs.

\section{Discussion}

\subsection{Physical models}

The above results suggest a picture in which BAL RLQs are in some
sense intermediate between BAL RQQs and non-BAL RLQs: BAL RLQs are
X-ray weak, but not to the same relative degree as are BAL RQQs, and
they can have harder \hbox{X-ray} spectra, but often have hardness ratios
consistent with those of non-BAL RLQs. A simple physical model could
also portray BAL RLQs as having \hbox{X-ray} characteristics of both BAL RQQs
(an outflowing BAL wind that is associated with an \hbox{X-ray} absorber) and
non-BAL RLQs (an unresolved \hbox{X-ray} emitting jet that contributes to the
total continuum). There are too many free parameters to constrain such
a model in detail, but some insight can be gained by making the
simplifying assumptions that the disk-corona emission in RLQs has the
same optical/UV-to-X-ray properties as are observed in RQQs [i.e.,
  that any systematic differences in the accretion structure of RLQs
  compared to RQQs do not produce dramatic changes in the $l_{\rm
    x}(l_{\rm uv})$ relation], and that the optical luminosity in RLQs
is dominated by disk-related emission with only a minimal jet-linked
contribution (certainly plausible for these broad-line RLQs, and often
inferred for even RLQs in which the radio and \hbox{X-ray} emission is
established as jet-dominated; e.g., Sambruna et al.~2006). Then the
disk-corona \hbox{X-ray} luminosity in RLQs may be calculated using the
$l_{\rm x}(l_{\rm uv})$ relation for RQQs, and any additional X-ray
luminosity may be ascribed to jet-linked emission.

The ratio of total RLQ \hbox{X-ray} luminosity to equivalent RQQ
(disk-corona) \hbox{X-ray} luminosity increases with increasing radio
luminosity (Figure 13a); this presumably reflects increasing jet
luminosity at both radio and \hbox{X-ray} frequencies with decreasing
inclination. The precise nature of the \hbox{X-ray} jet emission in RLQs (and
its dependence upon inclination) remains a matter of debate, although
it seems likely that two-zone models are required (e.g., Jester et
al.~2006), in which beamed jet-linked \hbox{X-ray} emission from the fast
spine dominates for objects viewed at low inclinations while a slower
(and less beamed) sheath could generate jet-linked \hbox{X-ray} emission
radiated in a more isotropic manner. We refrain from imposing a
particular jet model upon the data but quantify the observed increase
in the \hbox{X-ray} luminosity of RLQs relative to RQQs with increasing radio
luminosity through the trendline shown in Figure 13a. Non-BAL RLQs
with optical/UV and radio luminosities similar to those of the BAL
RLQs in our sample would have total \hbox{X-ray} luminosities greater than
the non-BAL RQQ equivalent by a typical multiplicative factor of
1.8--2.7, suggesting roughly equal contributions from disk-corona and
jet-linked \hbox{X-ray} emission.

If we further presume that the \hbox{X-ray} absorber in BAL RLQs has
characteristics similar to those found for BAL RQQs, then based on
Figure 10c the disk-corona emission ought to be reduced by a factor of
$\sim$10 in BAL RLQs relative to non-BAL RLQs. If the jet-linked X-ray
emission were also absorbed to a similar degree, the entire X-ray
continuum in BAL RLQs would be veiled as in BAL RQQs and the relative
X-ray luminosities and hardness ratios of BAL RLQs would agree better
with those of BAL RQQs, contrary to observation. However, it seems
that the jet must be partially covered by the BAL-linked X-ray
absorber in order to explain the difference between predicted and
observed jet-linked \hbox{X-ray} emission in BAL RLQs. Specifically, we find
that many BAL RLQs have jet-linked \hbox{X-ray} emission only \hbox{20--80\%}
of that expected, and further those BAL RLQs with hardness ratios
harder than 90\% of RLQs tend to have less jet-linked \hbox{X-ray} emission
than predicted (Figure~13b). We postulate that the BAL-linked X-ray
absorber is of sufficient size to cover some fraction of the X-ray
emitting jet in many BAL RLQs.

Alternative scenarios are possible; while we cannot rule them out,
they are difficult to motivate either physically or from the data. It
might be surmised that BAL RLQs are intrinsically \hbox{X-ray} weak relative
to RLQs and are also (typically) unabsorbed. The \hbox{X-ray} absorber is
thought to shield the BAL wind from overionization in disk-wind models
(e.g., Murray et al.~1995), so such a postulated lack of an X-ray
absorber could require BAL formation and acceleration to occur in a
manner distinct from that in BAL RQQs. One mechanism by which BAL RLQs
could be intrinsically \hbox{X-ray} weak would be if the presence of a BAL
outflow inhibited the production of small-scale jet-linked X-ray
emission. If the disk/corona system were relatively similar to that of
RQQs, then BAL RLQs should follow the non-BAL RQQ luminosity
correlations; however, BAL RLQs with lower radio luminosities have
X-ray luminosities less than those of non-BAL RQQs with comparable
optical/UV luminosities (Figures 9a and 13a). Another mechanism by
which BAL RLQs could be intrinsically \hbox{X-ray} weak would be for the
disk/corona system to be an inefficient emitter of X-rays. If the
small-scale \hbox{X-ray} emitting jet were relatively similar to that in
RLQs, then as the fractional contribution in RLQs from the disk/corona
would be expected to decrease at high radio luminosities, the
difference in \hbox{X-ray} luminosity between BAL RLQs and RLQs should
likewise decrease; however, the offset between BAL RLQs and RLQs
appears roughly constant (Figures 9c and 9d) across the two orders of
magnitude in radio luminosity spanned by our sample. In any event, it
seems most straight-forward to retain those fundamental features
firmly established as present in BAL RQQs or RLQs when interpreting
BAL RLQs, and the simple model associated with Figure 13 and described
above suffices to explain the current data naturally.

\subsection{BAL RLQ geometries and ages}

There have been suggestions that BAL outflows occur at low
inclinations (e.g., Brotherton et al.~2006; Zhou et al.~2006),
something difficult to explain from simple disk-wind models. Some of
the BAL RLQs in our sample have radio properties consistent with the
jet being pointed close to the line of sight, including compact
morphologies and flat radio spectra, although GPS sources with
sparsely sampled radio spectra can mimic such characteristics at a
range of inclinations. It does not seem likely that outflows in BAL
RLQs must always be polar, since some of the BAL RLQs in our sample
are steep-spectrum objects dominated by extended radio emission,
arguing against low inclinations for these objects. We can estimate
the inclinations of the core-dominated BAL RLQs using the
core-radio-to-optical luminosity ratio (essentially the radio
loudness, excluding lobe emission; Wills \& Brotherton 1995), and find
probable inclinations of $\sim$20$^{\circ}$ to $>40^{\circ}$, but this
method is insensitive to larger inclinations and may not apply to BAL
RLQs. The sample BAL RLQs do not tend to have large values of $R^{*}$
($\simgt$500--1000), with the two exceptions being a lobe-dominated
and a CSS RLQ, suggesting most are not low inclination sources
(cf. Figure 1 of Wills \& Brotherton 1995). 

The discovery of BALs in RLQs which are compact and have radio spectra
similar to those of presumed young GPS or CSS sources has led to
suggestions that BALs are associated with a quasar evolutionary
phase.\footnote{The relatively high fraction of low-ionization BAL
  quasars among dust-reddened quasars has also motivated the
  association of (at least low-ionization) BALs with young quasars
  (e.g., Urrutia et al.~2009).} Not all GPS or CSS sources display
BALs, and not all BAL RLQs are associated with young
sources;\footnote{As in our sample, BAL RLQs can be found in FR~IIs
  with large projected sizes, although the rarity of such objects is
  interpreted by Gregg et al.~(2006) as support for an evolutionary
  scenario.} if there is no inclination dependence to BALs then (as
assessed by Shankar et al.~2008) strictly evolutionary models require
problematic fine tuning of the various phases to match
observations. It is possible to compare directly the X-ray properties
of GPS and CSS sources to those of BAL RLQs to search for
similarities. We plot data from {\it Chandra\/} observations of GPS
and CSS radio galaxies and RLQs carried out by Siemiginowska et
al.~(2008) on the $l_{\rm x}(l_{\rm uv})$ and $l_{\rm x}(l_{\rm r})$
relations shown earlier (Figure 14). GPS/CSS sources are often X-ray
weak relative to RLQs of similar optical/UV luminosity, but they are
also often extremely radio-loud (see also the spectral energy
distribution plots from Siemiginowska et al.~2008) in a manner that
the (non-GPS, non-CSS) BAL RLQs are not. The CSS BAL RLQ
J104834.24+345724.9 is both radio luminous and \hbox{X-ray} weak, with
a hard X-ray spectrum atypical of GPS/CSS sources, as might be
expected for an object in both classes.

\section{Conclusions}

This work presents and discusses the \hbox{X-ray} properties of 21 BAL
RLQs observed with {\it Chandra}. The sample of BAL RLQs spans a wide
range of C~IV absorption properties, is dominated by high-ionization
BAL quasars, is restricted to definitively radio-loud quasars with
$R^{*}\simgt50$, and includes objects with both core-dominated and
lobe-dominated radio morphologies. We find the following results:

1. BAL RLQs are \hbox{X-ray} weak relative to non-BAL RLQs of similar
optical/UV luminosity, but not to as extreme a degree as are BAL RQQs
relative to comparable non-BAL RQQs. BAL RLQs are also \hbox{X-ray} weak, to
a lesser extent, relative to non-BAL RLQs of similar radio luminosity
or of both similar optical/UV and radio luminosities.

2. BAL RLQs do not show a strong correlation between \hbox{X-ray} spectral
hardness and \hbox{X-ray} weakness, as is observed in BAL RQQs, and do not
tend to have as extreme hardness ratios as can BAL RQQs.

3. The simplest model to explain our results is that the X-ray
continuum in BAL RLQs consists of both disk/corona and jet-linked
X-ray emission; absorption of the disk/corona emission alone typically
will neither reduce the observed \hbox{X-ray} luminosity nor harden the X-ray
spectrum of BAL RLQs to the same degree as in BAL RQQs.

4. Although jet-linked \hbox{X-ray} emission in BAL RLQs does not generally
appear to be absorbed to the same degree as is the \hbox{X-ray} continuum in
BAL RQQs, it does seem likely that the \hbox{X-ray} emitting small-scale jet
is partially covered in many BAL RLQs.

Microquasar observations have been interpreted to show that in the
soft state a radiatively-driven disk wind develops and becomes the
dominant channel for outflow of accreting material, quenching the jet
(Neilsen \& Lee~2009). Although the dearth of BALs in strongly
radio-loud objects suggests a similar mechanism may apply to quasars,
it is clear that jets and winds can coexist in at least some
RLQs. Further \hbox{X-ray} studies can help clarify the relationship between
jets and outflows in RLQs: snapshot {\it Chandra\/} observations of
additional BAL RLQs could permit more quantitative consideration of
various physical models, while deep {\it XMM-Newton\/} spectral
observations of the brightest BAL RLQs would help elucidate the
properties of the \hbox{X-ray} absorber and perhaps differentiate them from
those in BAL RQQs. 

\acknowledgments

We gratefully acknowledge the financial support of NASA grant SAO
SV4-74018 (G.~P.~G., Principal Investigator) and NASA LTSA grant
NAG5-13035 (B.~P.~M., W.~N.~B.). We thank the referee for useful
comments, Mike Eracleous for helpful discussions as well as assistance
with HET/LRS data reduction and analysis, Jianfeng Wu for technical
advice, and Chris Willot and Bob Becker/Rick White for providing us
with electronic spectra of 4C~+35.23 and FBQS 0256$-$0119,
respectively.

Funding for the SDSS and SDSS-II has been provided by the Alfred
P. Sloan Foundation, the Participating Institutions, the National
Science Foundation, the U.S. Department of Energy, the National
Aeronautics and Space Administration, the Japanese Monbukagakusho, the
Max Planck Society, and the Higher Education Funding Council for
England. The SDSS Web Site is http://www.sdss.org/.

The Hobby-Eberly Telescope (HET) is a joint project of the University
of Texas at Austin, the Pennsylvania State University, Stanford
University, Ludwig-Maximillians-Universit\"at M\"unchen, and
Georg-August-Universit\"at G\"ottingen. The HET is named in honor of
its principal benefactors, William P. Hobby and Robert E. Eberly. The
Marcario Low-Resolution Spectrograph is named for Mike Marcario of
High Lonesome Optics, who fabricated several optics for the instrument
but died before its completion; it is a joint project of the
Hobby-Eberly Telescope partnership and the Instituto de
Astronom\'{\i}a de la Universidad Nacional Aut\'onoma de M\'exico.

\clearpage

\begin{deluxetable}{p{90pt}llrrllrrr}
\tablecaption{{\it Chandra} and {\it HET} Observing Log}
\tabletypesize{\scriptsize}

\tablehead{\colhead{} & \multicolumn{4}{c}{{\it Chandra} Observations}
& & \multicolumn{4}{c}{{\it HET} Observations} \\
\colhead{Name (SDSS)} & \colhead{ObsID} & \colhead{Date} &
\colhead{Exp (s)} & \colhead{Counts\tablenotemark{a}} &  &
\colhead{Date} & \colhead{Exp (s)} &
\colhead{$\lambda/\Delta\lambda$\tablenotemark{b}} &
\colhead{S/N\tablenotemark{c}}}

\startdata

074610.50$+$230710.8\dotfill & 9160 & 2007 Dec 12 & 6954 & 67.9$_{  -8.2}^{+   9.3}$ &~~ & 2007 Dec 18 & 1200   &867    & 10      \\
083749.59$+$364145.4\dotfill & 9153 & 2007 Dec 23 & 6101 & 17.7$_{  -4.2}^{+   5.3}$ &~~ & 2007 Dec 18 & 1800   &867    & 20      \\
085641.58$+$424254.1\dotfill & 9156 & 2008 Feb 10 & 5968 & 17.0$_{  -4.1}^{+   5.2}$ &~~ & 2008 Feb 24 & 1500   &867    & 18      \\
092913.96$+$375742.9\dotfill & 9162 & 2007 Dec 28 & 3987 & 47.6$_{  -6.9}^{+   7.9}$ &~~ & \nodata     & \nodata&\nodata& \nodata \\
102258.41$+$123429.7\dotfill & 9154 & 2008 Apr 04 & 4976 & 24.1$_{  -4.9}^{+   6.0}$ &~~ & 2008 May 03 & 900    &300    & 26      \\
105416.51$+$512326.0\dotfill & 9163 & 2008 Jan 20 & 4979 & 36.7$_{  -6.0}^{+   7.1}$ &~~ & 2008 Feb 08 & 1500   &867    & 19      \\
112506.95$-$001647.6\dotfill & 9157 & 2008 Apr 29 & 5112 & 80.1$_{  -8.9}^{+  10.0}$ &~~ & \nodata     & \nodata&\nodata& \nodata \\
115944.82$+$011206.9\dotfill & 9158 & 2007 Feb 28 & 3706 &171.1$_{ -13.1}^{+  14.1}$ &~~ & 2008 Feb 13 & 900    &867    & 37      \\
123411.73$+$615832.6\dotfill & 9152 & 2008 Feb 22 & 6142 & 17.4$_{  -4.1}^{+   5.2}$ &~~ & 2008 Apr 23 & 1500   &867    & 29      \\
133701.39$-$024630.3\dotfill & 9159 & 2007 Dec 10 & 4692 & 23.3$_{  -4.8}^{+   5.9}$ &~~ & 2008 Feb 08 & 1500   &867    & 17      \\
141334.38$+$421201.7\dotfill & 9161 & 2008 Apr 03 & 6954 & 48.6$_{  -6.9}^{+   8.0}$ &~~ & 2008 May 08 & 1200   &650    & 36      \\
162453.47$+$375806.6\dotfill & 9155 & 2007 Nov 25 & 3994 & 18.3$_{  -4.2}^{+   5.4}$ &~~ & 2008 Feb 08 & 1300   &867    & 15      \\

\enddata

\tablecomments{All targets were observed with {\it Chandra} on-axis
with the ACIS-S array, using Very Faint mode. {\it HET} spectra were
obtained with the Low Resolution Spectrograph; 10/12 targets were able
to be observed.}

\tablenotetext{a}{Background-subtracted and aperture-corrected counts
in the 0.5--8 keV band. Errors are 1$\sigma$ (Poisson errors; Gehrels
1986). All snapshot targets are detected.}

\tablenotetext{b}{Most observations were conducted using the g2 grism
with a 1.5$''$ slit, providing a resolving power of 867.}

\tablenotetext{c}{Signal-to-noise of the continuum near observed-frame
6000~\AA.}

\end{deluxetable}

\begin{deluxetable}{lp{95pt}llrrrrr}
\tablecaption{{\it Chandra} Archival Sources}
\tabletypesize{\scriptsize}

\tablehead{\colhead{Name (SDSS)} & \colhead{Name (Other)} &
\colhead{ObsID} & \colhead{Date} & \colhead{Exp (s)} &
\colhead{$\theta$ ($'$)\tablenotemark{a}} &
\colhead{Counts\tablenotemark{b}} & \colhead{Sel\tablenotemark{c}} &
\colhead{Ref}}

\startdata

020022.01$-$084512.0  & FBQS J0200$-$0845\dotfill      & 3265 & 2002 Oct 02  & 17901& 9.4&   55.3$_{  -7.4}^{+   8.5}$ & S & 1 \\    
\nodata               & FBQS J0256$-$0119\dotfill      &~~850 & 1999 Dec 09  & 4456 & 0.0& 18.8$_{  -4.3}^{+   5.4}$    & L & 2 \\
081426.45$+$364713.5  & \dotfill                       & 3436 & 2002 Jan 31  & 9839 & 8.3&    8.8$_{  -2.9}^{+   4.1}$ & S & \nodata \\
081426.45$+$364713.5\tablenotemark{d}   & \dotfill     & 3437 & 2002 Feb 11  & 9933 & 8.3&    9.5$_{  -3.0}^{+   4.2}$ & S  &\nodata  \\
091951.29$+$005854.9  & \dotfill                       & 7056 & 2006 Jun 28  & 5080 & 4.3& $<   3.6$ & S & \nodata \\
100726.10$+$124856.2  & PG 1004$+$130\dotfill          & 5606 & 2005 Jan 05  & 41064& 0.0& 1851.4$_{ -43.0}^{+  44.0}$ & L & 3 \\
104834.24$+$345724.9  & 4C $+$35.23\dotfill            & 9320 & 2008 Jan 20  & 4658 & 0.0&    5.5$_{  -2.3}^{+   3.5}$ & L & 4 \\
122033.87$+$334312.0  & 3C 270.1\dotfill               & 2118 & 2002 Apr 03  & 3087 & 4.3&  177.4$_{ -13.3}^{+  14.3}$ & S & \nodata \\
131213.57$+$231958.6  & FBQS J131213.5$+$231958\dotfill&~~852 & 2000 May 19  & 4686 & 0.0&   58.8$_{  -7.6}^{+   8.7}$ & L & 2 \\
\nodata               & LBQS 2211$-$1915\dotfill       & 4836 & 2003 Nov 19  & 5889 & 0.0&   53.8$_{  -7.3}^{+   8.4}$ & L & 5 \\

\enddata

\tablenotetext{a}{Off-axis angle in arc-minutes; a value of 0.0
indicates an observation targeting that BAL RLQ.}

\tablenotetext{b}{Background-subtracted and aperture-corrected counts
in the 0.5--8 keV band. Errors are 1$\sigma$ (Poisson errors; Gehrels
1986), while limits for non-detections are at the 95\% confidence
level (Bayesian statistics; Kraft et al.~1991).}

\tablenotetext{c}{Selection method: S = BAL RLQ identified from
SDSS/FIRST data with serendipitous {\it Chandra} archival coverage,
while L~=~BAL RLQ identified from literature with targeted {\it Chandra}
archival coverage.}

\tablenotetext{d}{Since the two observations of 081426.45+364713.5 are
  of comparable quality, both are shown; these observations are
  stacked for later analysis. There are 18.3$^{+5.4}_{-4.2}$ net
  0.5--8~keV counts in the combined 19.8~ks exposure.}

\tablerefs{Prior analysis of {\it Chandra} data: (1) Gallagher et
  al.~(2005); (2) Brotherton et al.~(2005); Miller et al.~(2006); (4)
  PI Kunert-Bajraszewska; (5) Gallagher et al.~(2006).}

\end{deluxetable}

\begin{deluxetable}{p{120pt}rrrrrrrrll}
\tablecaption{Optical/UV Characteristics}
\tabletypesize{\scriptsize}

\tablehead{\colhead{Name (SDSS)} & \colhead{$z$} & \colhead{$m_{\rm
i}$} & \colhead{$M_{\rm
i}$} & \colhead{${\Delta}(g-i)$} & \colhead{BI} & \colhead{AI} &
\colhead{EW} & \colhead{$V_{\rm max}$} &
\colhead{Type\tablenotemark{a}} & \colhead{Ref}}

\startdata
\cutinhead{Snapshot BAL RLQs}
074610.50+230710.8\dotfill   & 2.093	  & 18.27 &$-$27.19&1.057&       \nodata &2955	 &      14.4\tablenotemark{b} &  5225  &Hi&1 \\
083749.59+364145.4\dotfill   & 3.416      & 18.55 &$-$28.01&0.614& 	4243.0  &6881	 &      34.6 &	11472  &HL&2 \\
085641.58+424254.1\dotfill   & 3.062      & 18.41 &$-$27.91&0.106& 	820.6 	&2660	 &       12.8& 	8071   &H&2 \\
092913.96+375742.9\dotfill   & 1.915      & 17.51 &$-$27.76&0.562& 	0.0 	&2630	 &      13.4 &	4166   &Hi&2 \\
102258.41+123429.7\dotfill   & 1.729      & 17.86 &$-$27.16&0.562& 	1008.7  &\nodata &	10.1 &	10180  &Hi&2 \\
105416.51+512326.0\dotfill   & 2.341      & 18.48 &$-$27.24&0.352& 	337.5 	&2177	 &       9.7 & 	5669   &H&2 \\
112506.95$-$001647.6\dotfill & 1.770      & 18.86 &$-$26.22&0.264& 	0.0 	&1743	 &        7.1&  3897   &Hi&2 \\
115944.82+011206.9\dotfill   & 2.000      & 16.96 &$-$28.40&0.412& 	0.0 	&2887	 &       12.5& 	4331   &Hi&2 \\
123411.73+615832.6\dotfill   & 1.946      & 18.39 &$-$26.90&0.442& 	4907.9  &10718	 &      32.4 &	20020  &Hi&2 \\
133701.39$-$024630.3\dotfill & 3.064      & 18.41 &$-$27.91&0.242& 	0.0 	&1657	 &       5.3 & 	4983   &H&2 \\
141334.38+421201.7\dotfill   & 2.817      & 18.24 &$-$27.89&0.516& 	0.0 	&2688	 &      15.0 &	4861   &H&2 \\
162453.47+375806.6\dotfill   & 3.381      & 18.15 &$-$28.38&0.281& 	2990.0 	&4156	 &       13.6\tablenotemark{b}&  28300 &H&3 \\
\cutinhead{Archival BAL RLQs}                         
020022.01$-$084512.0\dotfill & 1.943      & 18.28 &$-$27.01&0.484& 	2788.2  &4135	 &      19.9 &	14231  &Hi&2 \\
FBQS J0256$-$0119\dotfill    & 2.490 	  & 18.40 &$-$27.46&  0  & 	250.0	&\nodata &	12.9\tablenotemark{b}   & 	12400\tablenotemark{b}  &H&4 \\
081426.45+364713.5\dotfill   & 2.732      & 19.81 &$-$26.26&1.104& 	2438.40 &5289	 &      26.2 &	8300   &HLF&2 \\
091951.29+005854.9\dotfill   & 2.114      & 19.45 &$-$26.04&0.130& 	673.1 	&1268	 &       6.9 & 	22041  &nHi&2 \\
100726.10+124856.2\dotfill   & 0.241	  & 15.20 &$-$25.09&$-$0.45& 	850.0	&\nodata &	14.7\tablenotemark{b} & 	10000  &Hi&5 \\  
104834.24+345724.9\dotfill   & 1.594	  & 20.17 &$-$24.66& 0.88& 	\nodata	&\nodata &	26.2\tablenotemark{b} &  14300\tablenotemark{b}  &Hi&6 \\
122033.87+334312.0\dotfill   & 1.532      & 18.10 &$-$26.64&0.398& 	52.5 	&\nodata &	6.5  &	5266   &HL&2 \\
131213.57+231958.6\dotfill   & 1.508	  & 17.11 &$-$27.59&0.084& 	1400.0  &\nodata &	15.6\tablenotemark{b} & 	25000  &Hi&7 \\
LBQS 2211$-$1915\dotfill     & 1.952	  & 17.34 &$-$27.96& 0.32& 	27.0	&\nodata &	7.8  &	11544  &Hi&8,9 \\
		   
\enddata

\tablecomments{The $k$-correction for $M_{\rm i}$ assumes a power-law
  continuum with spectral index ${\alpha}_{\nu}=-0.5$. The absorption
  properties refer to C~IV measurements. BI, EW, and $V_{\rm max}$
  values are primarily from the listed reference, chiefly Gibson et
  al.~(2009), while AI values are taken from Trump et al.~(2006) where
  available. The units for BI, AI, and $V_{\rm max}$ are km s$^{-1}$;
  EW is in \AA.}

\tablenotetext{a}{BAL type following Trump et al.~(2006): Hi = HiBAL
  (no Mg~II absorption in spectrum); HLF = FeLoBAL (C~IV BAL, Fe~II or
  Fe~III absorption in spectrum); HL = HiBAL with some low-ionization
  absorption; H = HiBAL lacking spectral coverage of Mg~II; n =
  relatively narrow absorption. Type is taken from Trump et al.~(2006)
  for all quasars with reported AI measurements; the remainder are
  classified based on our examination of the SDSS spectrum where
  available or else by the listed reference.}

\tablenotetext{b}{EW or $V_{\rm max}$ value measured by us.}

\tablerefs{(1) Trump et al.~(2006); (2) Gibson et al.~(2009); Benn et
al.~(2005); (4) Becker et al.~(2001); (5) Wills et al.~(1999); (6)
Willott et al.~(2002); (7) Becker et al.~(2000); (8) Weymann et
al.~(1991); (9) Gallagher et al.~(2006).}

\end{deluxetable}

\begin{deluxetable}{p{100pt}lrrrrrrr}
\tablecaption{Radio Characteristics}
\tabletypesize{\scriptsize}

\tablehead{ & & \multicolumn{4}{c}{Flux-Density
    Measurements\tablenotemark{a}} & \colhead{} &
  \multicolumn{2}{c}{Spectral Indices\tablenotemark{b}}
  \\ \colhead{Name (SDSS)} & \colhead{Type\tablenotemark{c}} &
  \colhead{$\leq$365 MHz} & \colhead{1.4 GHz (F)} & \colhead{1.4 GHz
    (N)} & \colhead{4.85 GHz} & & \colhead{${\alpha}_{\rm low}$} &
  \colhead{${\alpha}_{\rm high}$}}

\startdata
\cutinhead{Snapshot BAL RLQs}
074610.50+230710.8\dotfill   &P& \nodata        &23.68   &	22.3$\pm$0.8	&	27$\pm$4 G  	 &~~&	\nodata&	+0.11   \\
083749.59+364145.4\dotfill   &P& \nodata        &27.10   &       \nodata&\nodata&~~&	\nodata&	$-$0.43 \\
085641.58+424254.1\dotfill   &P& \nodata        &19.99   &	20.2$\pm$0.7	&	29$\pm$4 G  	 &~~&	\nodata&	+0.30	\\
092913.96+375742.9\dotfill   &P& 94 W	        &43.43   &	42.9$\pm$1.3	&	24$\pm$4 G  	 &~~&	$-$0.53&	$-$0.48	\\
102258.41+123429.7\dotfill   &D& 448$\pm$51 T   &118.65\tablenotemark{d}  &	126.1$\pm$3.8	&	44$\pm$6 G  	 &~~&	$-$0.99&	$-$0.80	\\
105416.51+512326.0\dotfill   &P& 44 W	        &33.88   &	35.6$\pm$1.1	&	22$\pm$4 G  	 &~~&	$-$0.18&	$-$0.35	\\
112506.95$-$001647.6\dotfill &D& 890$\pm$130 V  &65.47\tablenotemark{d}   &	74.0$\pm$2.7	&	\nodata	    	 &~~&	$-$0.89&	\nodata	\\
115944.82+011206.9\dotfill   &P& 887$\pm$30 T   &268.48  &	275.6$\pm$8.3   &       137.8$\pm$1.7 M	 &~~&	$-$0.89&	$-$0.54	\\
123411.73+615832.6\dotfill   &P& \nodata	&23.96   &	22.7$\pm$0.8	&         13.2 C	 &~~&	\nodata&	$-$0.48 \\	
133701.39$-$024630.3\dotfill &P& \nodata	&44.82   &	45.1$\pm$1.4	&         39.4 C   	 &~~&	\nodata&	$-$0.10	\\
141334.38+421201.7\dotfill   &P& 22 W	        &18.74   &	16.8$\pm$0.6	&       8.8$\pm$0.7 M	 &~~&	$-$0.11&	$-$0.61	\\
162453.47+375806.6\dotfill   &P& 72 W	        &56.44   &	55.6$\pm$1.7    &       23.3$\pm$1.1 B	 &~~&	$-$0.17&	$-$0.71 \\
\cutinhead{Archival BAL RLQs}
020022.01$-$084512.0\dotfill &P& \nodata	&7.34    &	8.0$\pm$0.5	&	\nodata	    	 &~~&	\nodata&	\nodata	\\
FBQS J0256$-$0119\dotfill    &P& \nodata	&27.56   &	22.3$\pm$0.8    &       12.0$\pm$0.5 M	 &~~&	\nodata&	$-$0.67	\\
081426.45+364713.5\dotfill   &P& \nodata	&2.98    &	3.8$\pm$0.4	&	\nodata	    	 &~~&	\nodata&	\nodata	\\
091951.29+005854.9\dotfill   &P& \nodata	&2.78    &	2.6$\pm$0.5	&	\nodata	    	 &~~&	\nodata&	\nodata	\\
100726.10+124856.2\dotfill   &D& 2740 P	        &\nodata\tablenotemark{d} &	1216.1$\pm$29.6\tablenotemark{d}&	415$\pm$37 G	 &~~&	$-$0.66&	$-$0.87	\\
104834.24+345724.9\dotfill   &P& 2437$\pm$29 T  &1050.97 &       1034.4$\pm$31.0&	439$\pm$39 G	 &~~&	$-$0.63&	$-$0.70	\\
122033.87+334312.0\dotfill   &D& 9742$\pm$134 T &2819.00\tablenotemark{d} &	2845.9$\pm$85.4	&	842$\pm$75 G	 &~~&	$-$0.92&	$-$0.97	\\
131213.57+231958.6\dotfill   &P& \nodata	&44.12   &	46.5$\pm$1.4    &       25.7$\pm$0.6 M	 &~~&	\nodata&	$-$0.43	\\
LBQS 2211$-$1915\dotfill     &P&\nodata        &\nodata &        64.0$\pm$2.0	&	\nodata	    	 &~~&	\nodata&	\nodata	\\

\enddata

\tablenotetext{a}{All flux density measurements are in mJy, taken from
the following sources: B = Benn et al.~(2005); C = archival VLA C-band
imaging; F = FIRST: Faint Images of the Radio Sky at Twenty cm,
integrated flux, RMS errors are $\simeq$0.15~mJy beam$^{-1}$ (White et
al.~1997); G = Green Bank 6-cm survey (Gregory et al.~1996); M =
Montenegro-Montes et al.~(2008); N = NVSS: NRAO VLA Sky Survey (Condon
et al.~1998); P = Parkes Catalogue 1990, 408 MHz; T = Texas Survey of
Radio Sources at 365 MHz (Douglas et al.~1996); V = VLA Low-Frequency
Sky Survey, 74 MHz (Perley et al.~2006); W = Westerbork Northern Sky
Survey, 326 MHz, RMS errors are $\simeq$4~mJy beam$^{-1}$ (Rengelink
et al.~1997).}

\tablenotetext{b}{Radio spectral indices include extended emission
  components, are given as $S_{\rm r}\propto{\nu}^{{\alpha}_{\rm r}}$,
  and use FIRST measurements where available (else NVSS measurements);
  the quantities ${\alpha}_{\rm low}$ and ${\alpha}_{\rm high}$ are
  calculated from the flux densities presented in the columns labeled
  $\leq$365 MHz and 4.85 GHz, respectively, in addition to the 1.4~GHz
  data. J083749.59+364145.4 has a bright unrelated radio source
  $\sim$50$''$ North of the core that contaminates low-resolution
  maps; the spectral index is from a high-resolution 8.45 GHz flux
  density measurement from Montenegro-Montes et al.~(2008).}

\tablenotetext{c}{Radio morphology: P = point source, D = double
  (lobes summed for flux measurements). See $\S3.3$ for comments.}

\tablenotetext{d}{Extended emission: J102258.41+123429.7 has two FIRST
  components offset from the SDSS position by 0.058$'$ and 0.174$'$,
  with integrated fluxes of 93.98 and 24.67 mJy, respectively;
  J112506.95$-$001647.6 has two FIRST components offset from the SDSS
  position by 0.099$'$ and 0.201$'$, with integrated fluxes of 55.36
  and 10.11 mJy, respectively; J100726.10+124856.2 (PG~1004+130) is
  over-resolved by FIRST, but has two NVSS components offset from the
  SDSS position by 0.434$'$ and 1.010$'$, with fluxes of 656.4 and
  559.7 mJy, respectively; J122033.87+334312.0 (3C 270.1) has two
  FIRST components offset from the SDSS position by 0.067$'$ and
  0.073$'$, with integrated fluxes of 2096.61 and 722.39 mJy,
  respectively.}

\end{deluxetable}

\begin{deluxetable}{p{80pt}rrrrcrrrrcrrrrr}
\tablecaption{X-ray Counts, Luminosities, and Properties of BAL RLQs}
\tabletypesize{\scriptsize} 
\setlength{\tabcolsep}{0.03in} 

\tablehead{\colhead{} & \multicolumn{4}{c}{X-ray
Counts\tablenotemark{a}} & &
\multicolumn{3}{c}{Luminosities\tablenotemark{b}} & &
\multicolumn{5}{c}{Derived Properties\tablenotemark{c}} \\
\colhead{Name (SDSS)} & \colhead{Soft} & \colhead{Hard} &
\colhead{$HR$} & \colhead{Rate} & & \colhead{$l_{\rm r}$} &
\colhead{$l_{\rm uv}$} & \colhead{$l_{\rm x}$} & &
\colhead{$R^{*}$} & \colhead{${\alpha}_{\rm ox}$} &
\colhead{${\Delta}l_{\rm x,uv}$} & \colhead{${\Delta}l_{\rm
x,S}$} & \colhead{${\Delta}l_{\rm x,H}$}}

\startdata
\cutinhead{Snapshot BAL RLQs}
074610.50+230710.8 	 &  49.4$_{  -7.0}^{+   8.1}$ &   17.6$_{  -4.2}^{+   5.3}$ & $-0.47_{-0.10}^{+ 0.11}$ &    9.8$_{  -1.2}^{+   1.3}$ &~~ &   33.36	 &31.15 &    27.01  &~~ &2.21  &$-$1.59  &$-$0.37  &$-$0.06  &$-$0.15 \\      
083749.59+364145.4 	 &  13.6$_{  -3.6}^{+   4.8}$ & $<   3.5$                 & $<-0.59$                &    2.9$_{  -0.7}^{+   0.9}$ &~~ &      33.85  &31.53 &    26.91      &~~  &2.32  &$-$1.77  &$-$0.81  &$-$0.52  &$<$$-$0.7 \\      
085641.58+424254.1 	 &  11.5$_{  -3.3}^{+   4.5}$ &    6.1$_{  -2.4}^{+   3.6}$ & $-0.31_{-0.19}^{+ 0.24}$ &    2.8$_{  -0.7}^{+   0.9}$ &~~ &   33.63  &31.43 &    26.79         &~~  &2.20  &$-$1.78  &$-$0.84  &$-$0.55  &$-$0.45 \\      
092913.96+375742.9 	 &  26.3$_{  -5.1}^{+   6.2}$ &   21.0$_{  -4.5}^{+   5.6}$ & $-0.11_{-0.14}^{+ 0.14}$ &   11.9$_{  -1.7}^{+   2.0}$ &~~ &   33.54  &31.40 &    26.99         &~~  &2.15  &$-$1.69  &$-$0.61  &$-$0.40  &$-$0.11 \\      
102258.41+123429.7 	 &  15.8$_{  -3.9}^{+   5.1}$ &    7.7$_{  -2.7}^{+   3.9}$ & $-0.34_{-0.17}^{+ 0.20}$ &    4.9$_{  -1.0}^{+   1.2}$ &~~ &   33.83  &31.13 &    26.52         &~~  &2.71  &$-$1.77  &$-$0.84  &$-$0.78  &$-$0.72 \\     
105416.51+512326.0 	 &  26.0$_{  -5.1}^{+   6.2}$ &   11.0$_{  -3.3}^{+   4.4}$ & $-0.41_{-0.13}^{+ 0.16}$ &    7.4$_{  -1.2}^{+   1.4}$ &~~ &   33.62  &31.27 &    26.96         &~~  &2.35  &$-$1.65  &$-$0.53  &$-$0.29  &$-$0.27 \\      
112506.95$-$001647.6 	 &  59.5$_{  -7.7}^{+   8.8}$ &   24.3$_{  -4.9}^{+   6.0}$ & $-0.42_{-0.09}^{+ 0.10}$ &   15.7$_{  -1.7}^{+   2.0}$ &~~ &   33.60  &30.74 &    27.05         &~~  &2.86  &$-$1.42  &$+$0.04  &$+$0.08  &$+$0.05 \\      
115944.82+011206.9 	 &  93.1$_{  -9.6}^{+  10.7}$ &   80.1$_{  -8.9}^{+  10.0}$ & $-0.08_{-0.08}^{+ 0.08}$ &   46.2$_{  -3.5}^{+   3.8}$ &~~ &   34.38  &31.67 &    27.62         &~~  &2.70  &$-$1.56  &$-$0.23  &$-$0.24  &$+$0.07 \\      
123411.73+615832.6 	 &  11.6$_{  -3.3}^{+   4.5}$ &    5.4$_{  -2.3}^{+   3.5}$ & $-0.36_{-0.19}^{+ 0.24}$ &    2.8$_{  -0.7}^{+   0.9}$ &~~ &   33.30  &31.06 &    26.38         &~~  &2.24  &$-$1.80  &$-$0.92  &$-$0.67  &$-$0.61 \\       
133701.39$-$024630.3 	 &  10.5$_{  -3.2}^{+   4.3}$ &   12.2$_{  -3.4}^{+   4.6}$ & $ 0.07_{-0.20}^{+ 0.20}$ &    5.0$_{  -1.0}^{+   1.3}$ &~~ &   33.98  &31.48 &    27.04         &~~  &2.50  &$-$1.70  &$-$0.63  &$-$0.65  &$-$0.21 \\      
141334.38+421201.7 	 &  34.7$_{  -5.9}^{+   6.9}$ &   16.5$_{  -4.0}^{+   5.1}$ & $-0.35_{-0.12}^{+ 0.14}$ &    7.0$_{  -1.0}^{+   1.2}$ &~~ &   33.53  &31.56 &    27.10         &~~  &1.97  &$-$1.71  &$-$0.65  &$-$0.24  &$-$0.18 \\      
162453.47+375806.6 	 &  13.4$_{  -3.6}^{+   4.7}$ &    5.5$_{  -2.3}^{+   3.5}$ & $-0.42_{-0.17}^{+ 0.23}$ &    4.6$_{  -1.1}^{+   1.3}$ &~~ &   34.17  &31.74 &    27.08         &~~  &2.43  &$-$1.79  &$-$0.83  &$-$0.61  &$-$0.60 \\      
\cutinhead{Archival BAL RLQs}                                     
020022.01$-$084512.0 	 & 38.4$_{  -6.2}^{+   7.2}$  &   14.9$_{  -3.8}^{+   4.9}$ & $-0.58_{-0.11}^{+ 0.13}$ &    3.1$_{  -0.4}^{+   0.5}$ &~~ &   32.79  &31.11 &    26.52         &~~  &1.68  &$-$1.76  &$-$0.82  &$-$0.31  &$-$0.36 \\      
FBQS J0256$-$0119     	 & 15.4$_{  -3.9}^{+   5.0}$  &    4.4$_{  -2.0}^{+   3.2}$ & $-0.56_{-0.15}^{+ 0.21}$ &    4.2$_{  -1.0}^{+   1.2}$ &~~ &   33.50	 &31.51 &    26.77  &~~ &1.99  &$-$1.82  &$-$0.93  &$-$0.51  &$-$0.59 \\      
081426.45+364713.5 	 & 10.2$_{  -3.1}^{+   4.3}$  &    7.5$_{  -2.7}^{+   3.8}$ & $-0.29_{-0.21}^{+ 0.23}$ &    0.9$_{  -0.2}^{+   0.3}$ &~~ &   32.70  &30.71 &    26.33          &~~ &1.99  &$-$1.68  &$-$0.65  &$-$0.38  &$-$0.17 \\      
091951.29+005854.9 	 &$<   3.3$                 & $<   3.5$                 & \nodata                 & $<   0.7$                 &~~ &          32.44  &30.74 &    $<$26.0   &~~ &1.71  &$<$$-$1.8  &$<$$-$1.0  &$<$$-$0.3  &$<$$-$0.0 \\     
100726.10+124856.2 	 &1107$_{ -33.3}^{+  34.3}$ &  794$_{ -28.2}^{+  29.2}$ & $-0.16_{-0.02}^{+ 0.02}$ &   45.1$_{  -1.0}^{+   1.1}$ &~~ &   32.86	 &30.52 &    25.66 &~~  &2.34  &$-$1.87  &$-$1.15  &$-$0.99  &$-$0.77 \\      
104834.24+345724.9       &$<   3.2$                 &    4.4$_{  -2.0}^{+   3.2}$ & $> 0.16$                &    1.2$_{  -0.5}^{+   0.7}$ &~~ &      34.76	 &30.48 &    25.82  &~~  &4.28  &$-$1.79  &$-$0.95  &$<$$-$1.6  &$-$1.10 \\      
122033.87+334312.0 	 &134.1$_{ -11.6}^{+  12.6}$  &   47.6$_{  -6.9}^{+   7.9}$ & $-0.48_{-0.10}^{+ 0.10}$ &   57.5$_{  -4.3}^{+   4.6}$ &~~ &   35.07  &30.89 &    27.41          &~~ &4.19  &$-$1.34  &$+$0.27  &$-$0.26  &$-$0.20 \\      
131213.57+231958.6 	 & 48.4$_{  -6.9}^{+   8.0}$  &   11.0$_{  -3.3}^{+   4.4}$ & $-0.63_{-0.09}^{+ 0.11}$ &   12.5$_{  -1.6}^{+   1.9}$ &~~ &   33.30	 &31.42 &    26.74  &~~  &1.88  &$-$1.80  &$-$0.88  &$-$0.42  &$-$0.55 \\      
LBQS 2211$-$1915     	 & 38.4$_{  -6.2}^{+   7.2}$  &   16.6$_{  -4.0}^{+   5.1}$ & $-0.40_{-0.11}^{+ 0.13}$ &    9.1$_{  -1.2}^{+   1.4}$ &~~ &   33.78	 &31.67 &    26.87  &~~  &2.11  &$-$1.84  &$-$0.98  &$-$0.66  &$-$0.54 \\      

\enddata

\tablenotetext{a}{The soft and hard bands are 0.5--2~keV and 2--8~keV,
  respectively; errors/limits are as in Tables 1 and 2. The hardness
  ratio is $HR = (H-S)/(H+S)$, where $S$ ($H$) is the soft (hard) band
  counts; errors are 1$\sigma$. The $HR$ values for
  J020022.01$-$084512.0 and J081426.45+364713.5, observed with ACIS-I,
  have been adjusted by subtracting 0.14 to enable direct comparison
  to the ACIS-S $HR$ values. Rate is counts~ks$^{-1}$ in the
  0.5--8~keV band.}

\tablenotetext{b}{These monochromatic luminosities have units of
  log~ergs~s$^{-1}$~Hz$^{-1}$, at rest-frame frequencies of 5~GHz,
  2500~\AA, and 2~keV for $l_{\rm r}$, $l_{\rm uv}$, and $l_{\rm x}$,
  respectively. $l_{\rm x}$ is calculated from the 0.5--8~keV count
  rates for a power-law of $\Gamma=1.5$ and is corrected for Galactic
  absorption.}

\tablenotetext{c}{The radio loudness (in log units) is \hbox{$R^{*} =
l_{\rm r}-l_{\rm uv}$} and the optical/UV-to-X-ray spectral slope is
\hbox{${\alpha}_{\rm ox} = 0.3838\times(l_{\rm x}-l_{\rm uv})$}. The
relative \hbox{X-ray} luminosities are \hbox{${\Delta}l_{\rm x,uv} = l_{\rm
x} - (0.905{\times}l_{\rm uv}-0.813)$} and \hbox{${\Delta}l_{\rm x,S/H}
= l_{\rm x,S/H} - (0.472{\times}l_{\rm uv} + 0.413{\times}l_{\rm r} -
1.392)$}, where $l_{\rm x,S}$ and $l_{\rm x,H}$ are 2~keV X-ray
luminosities calculated from the 0.5--2~keV and 2--8~keV count rates,
respectively.}

\end{deluxetable}

\clearpage

\begin{figure}
\includegraphics[scale=0.9]{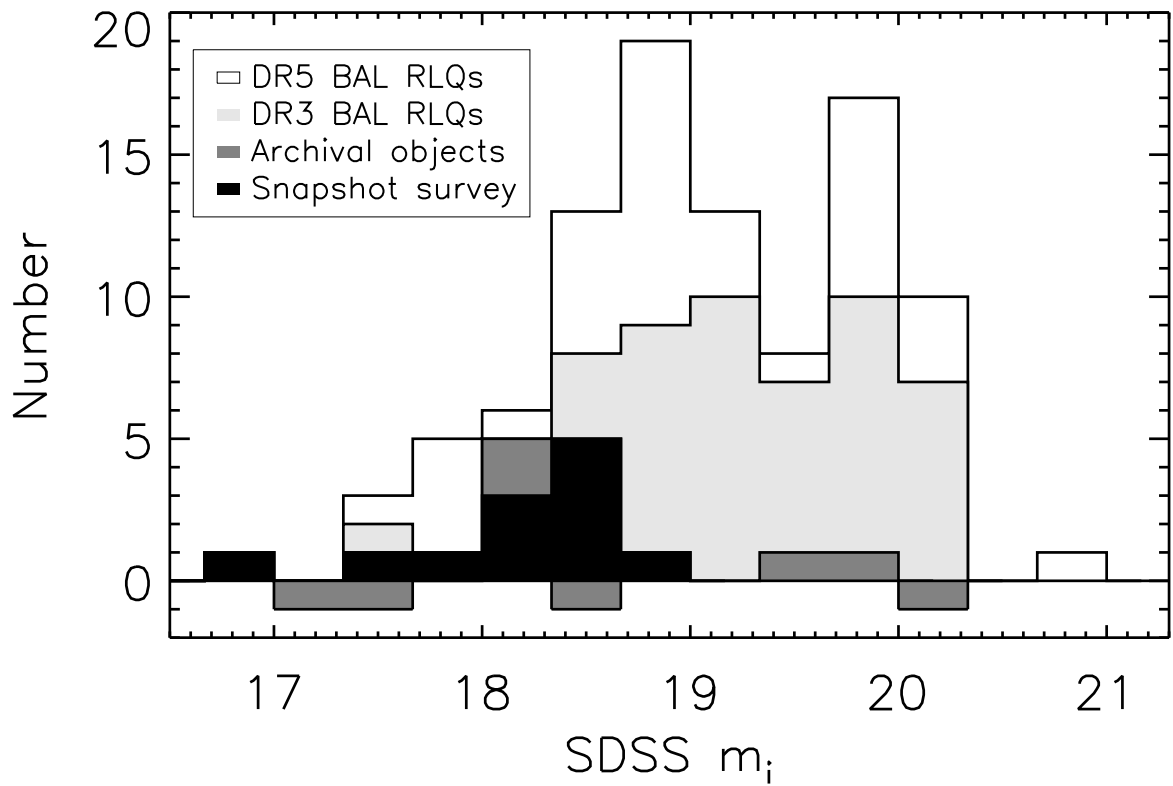}
\figcaption{\small Histogram showing the SDSS $m_{\rm i}$ distribution
for radio-loud broad absorption line quasars satisfying
$R^{*}\simgt100$ and EW$_{\rm CIV}>5$\AA. The light gray shaded
area indicates the subset of BAL RLQs that had spectra available in
SDSS DR3 or earlier, and the filled area shows those objects selected
for the snapshot survey (black, 12 objects) or possessing archival
{\it Chandra} coverage (dark gray, 8 objects shown; PG~1004+130 has
$m_{\rm i}=15.2$). Archival objects selected from the literature are
plotted on a negative scale.}
\end{figure}

\begin{figure}
\includegraphics[scale=0.85]{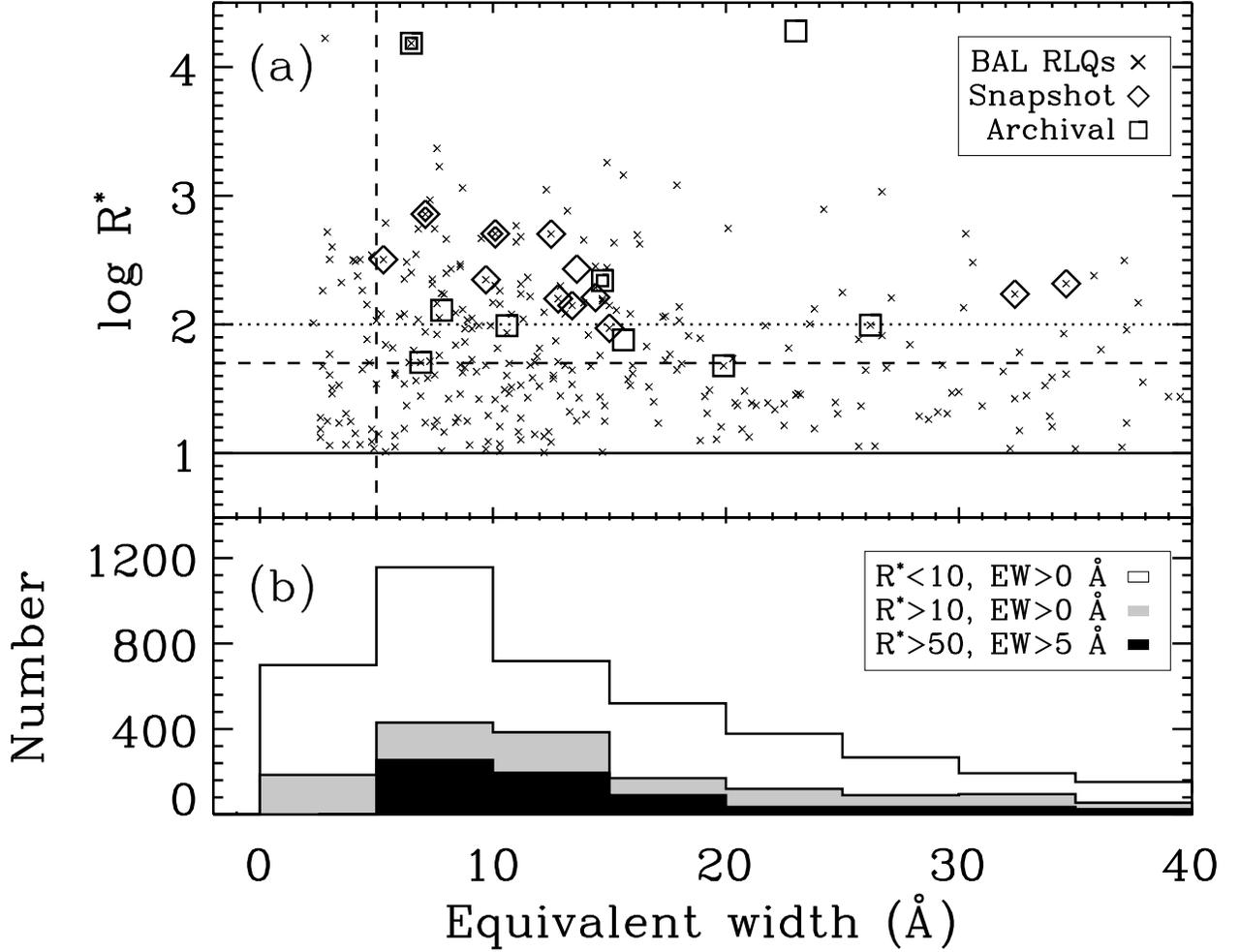} \figcaption{\small (a) Radio
  loudness plotted versus broad absorption line strength
  (parameterized by C~IV EW). The {\it Chandra} snapshot BAL RLQs are
  shown as diamonds and the archival BAL RLQs as squares (nested
  symbols are lobe-dominated BAL RLQs). The solid line marks the
  \hbox{$R^{*}>10~ (\log{R^{*}}>1)$} boundary, below which quasars are
  defined to be radio-quiet. The dashed lines show the selection
  criteria for the archival sample of BAL RLQs, which were required to
  be definitively radio-loud ($R^{*}\simgt50$) and show strong broad
  absorption lines ($EW>5$\AA). The dotted line shows the more
  restrictive criteria of $R^{*}\simgt100$ that was used to select the
  snapshot sample. As reported by previous authors it is rare for
  quasars to be simultaneously strongly absorbed and strongly
  radio-loud. (b) Plot of the distribution of C~IV EW for objects with
  $R^{*}<10$ (open histogram), objects with $R^{*}>10$ (gray
  histogram), and objects with $R^{*}>50$ and $EW>5$\AA~(black
  histogram). Numbers for objects with $R^{*}>10$ (the gray and black
  histograms) have been multiplied by 5 for clarity.}
\end{figure}

\begin{figure}
\includegraphics[scale=0.8]{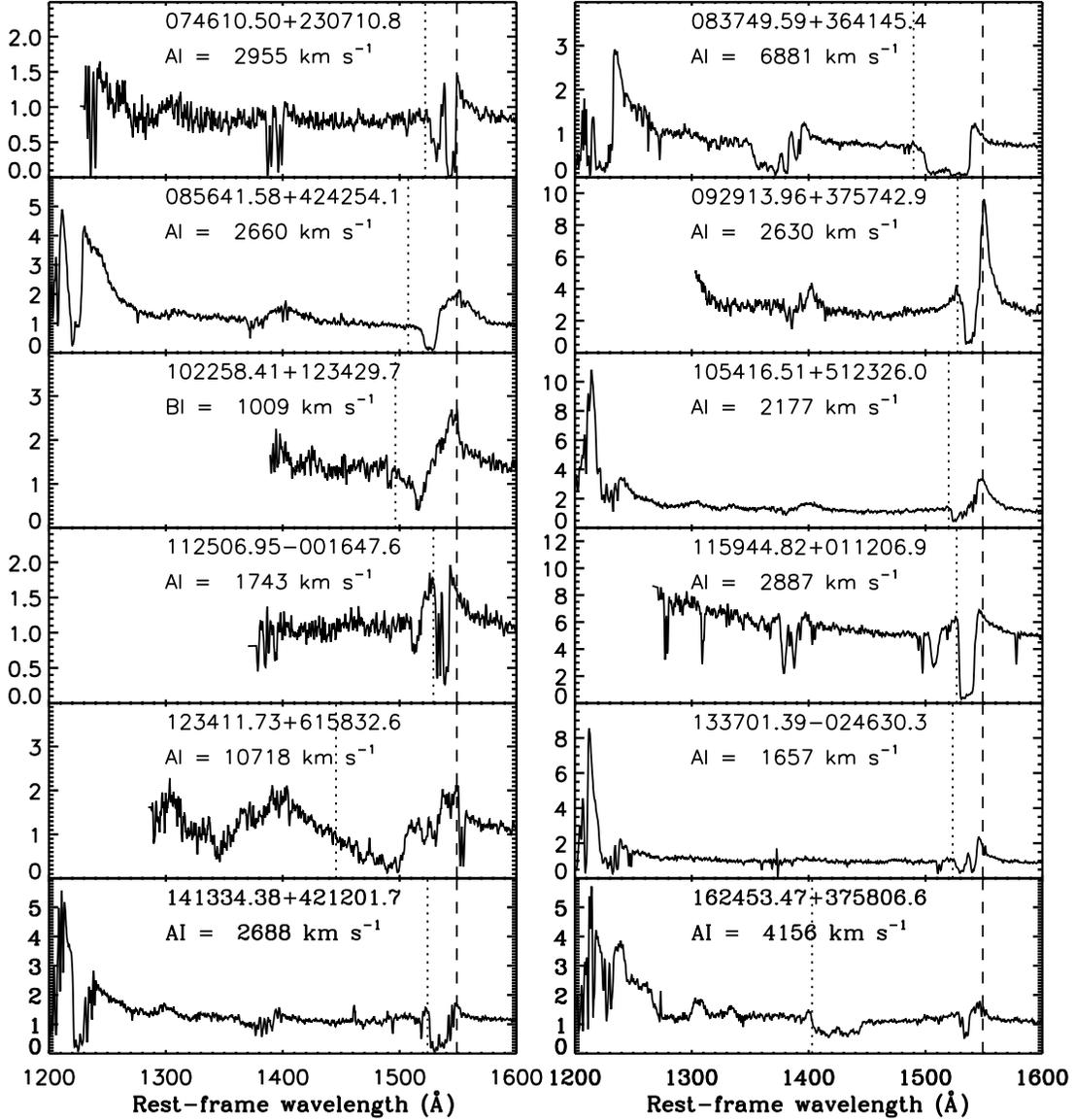}
\figcaption{\small SDSS spectra for the snapshot sample, plotted with
rest-frame wavelengths and showing the L$\alpha$ to C~IV BAL
region. The dashed line in each panel is at 1549~\AA, or zero
velocity. The dotted line indicates the maximum outflow velocity,
primarily taken from Gibson et al.~(2009). Flux is given in units of
10$^{-16}$ erg~cm$^{-2}$~s$^{-1}$~\AA$^{-1}$. The sample includes
objects covering a range of BAL absorption strengths and outflow
velocities. Each panel is labeled with the SDSS DR5 name as well as
the absorption index (a measure of BAL strength), primarily taken from
Trump et al.~(2006).}
\end{figure}

\begin{figure}
\includegraphics[scale=0.75]{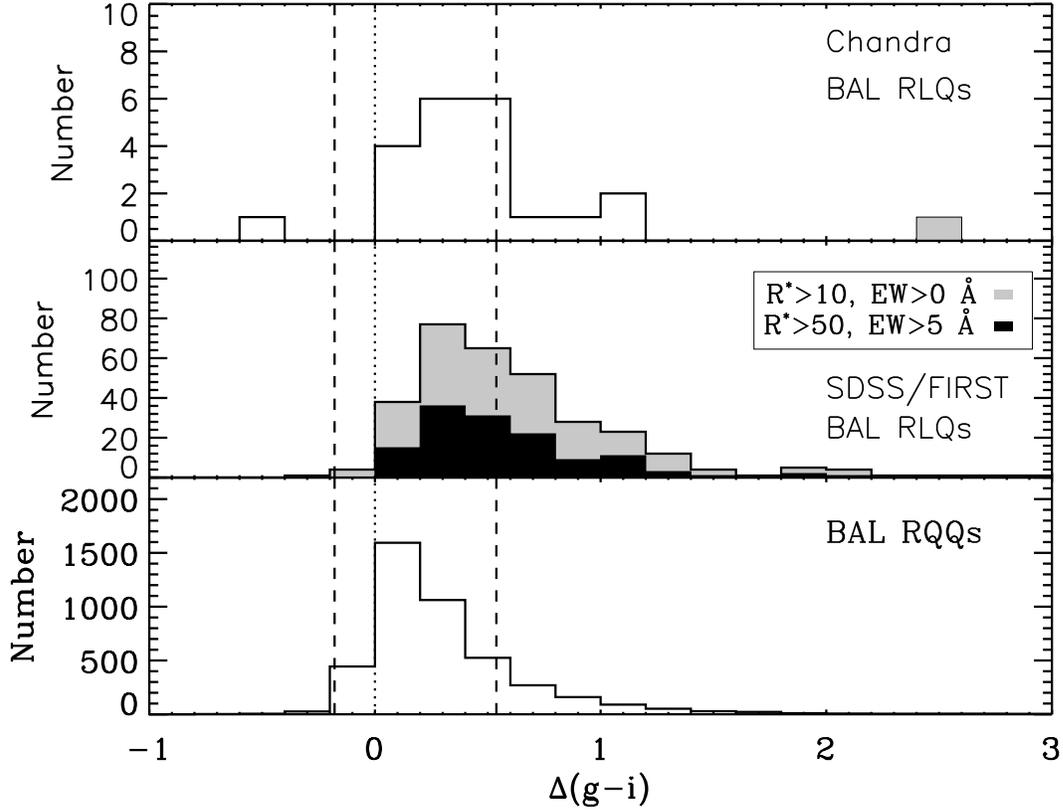} 
\figcaption{\small Relative color ${\Delta}(g-i)$, calculated by
  taking the measured $(g-i)$ for a given object and subtracting the
  $(g-i)$ that is typical for quasars at that redshift (positive
  values correspond to redder objects). The top panel shows the
  snapshot and archival sample of BAL RLQs with {\it Chandra} coverage
  (the object shaded in gray is the strongly reddened object FBQS
  J1556+3517, marked for comparison but not included in the archival
  sample). The middle panel shows SDSS/FIRST BAL RLQs, and the bottom
  panel shows BAL RQQs. The dashed lines enclose 90\% of SDSS DR5
  quasars. While BAL quasars tend to be redder than non-BAL quasars,
  the BAL RLQs studied with {\it Chandra} are representative of BAL
  RLQs in general and do not show excessive intrinsic reddening that
  could significantly elevate radio-loudness values.}
\end{figure}

\begin{figure}
\includegraphics[scale=0.85]{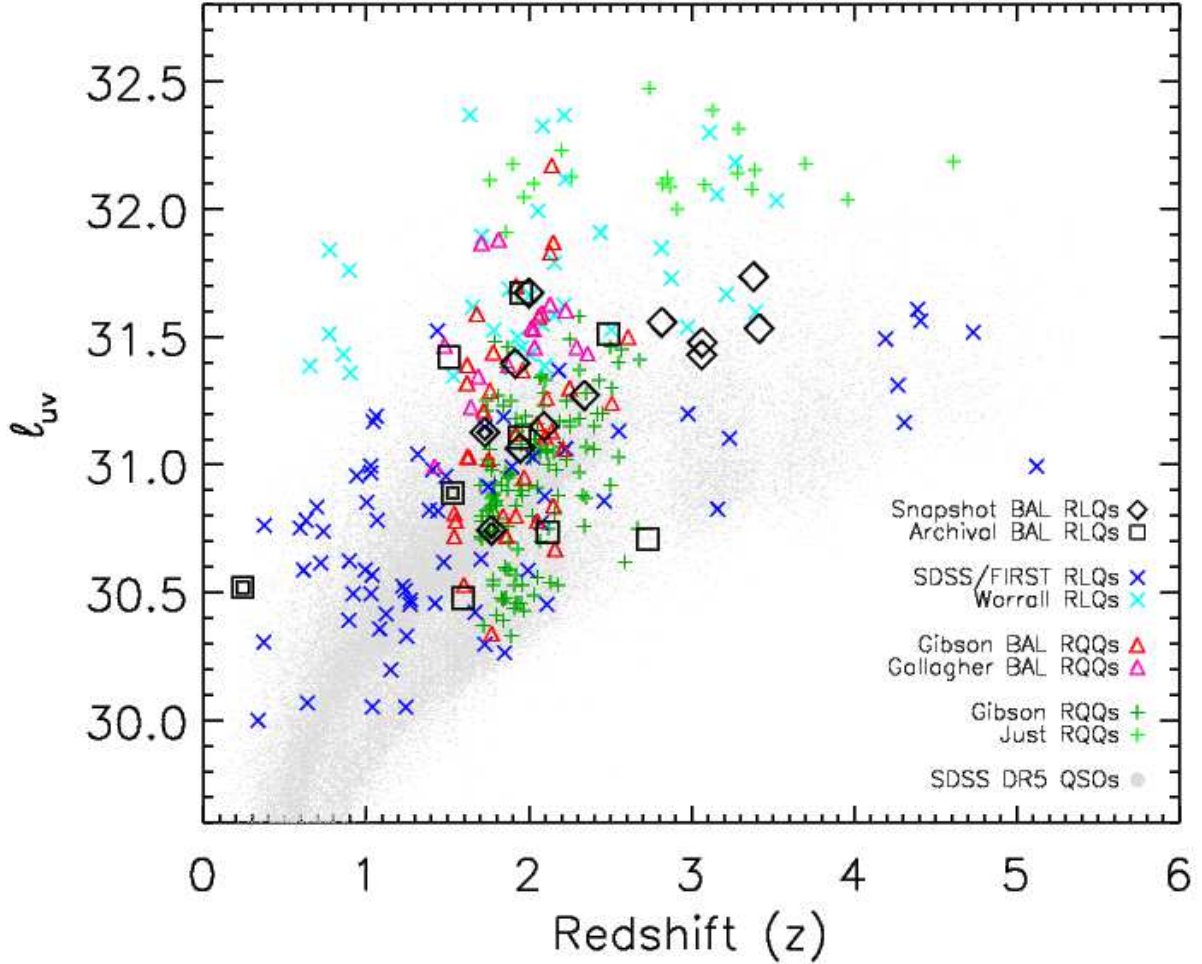} \figcaption{\small Optical
  luminosity $l_{\rm uv}$ in units of log~ergs~s$^{-1}$~Hz$^{-1}$
  calculated at rest-frame 2500~\AA~for our sample of BAL RLQs (nested
  symbols are lobe-dominated BAL RLQs) and for comparison samples of
  non-BAL RLQs, BAL RQQs, and non-BAL RQQs, plotted versus
  redshift. The RLQ comparison sample is constructed from
  SDSS/FIRST/{\it Chandra} data, supplemented with some particularly
  luminous RLQs observed by {\it Einstein} (Worrall et al.~1987). The
  BAL RQQ comparison sample is taken from Gibson et al.~(2009) and is
  supplemented with non-SDSS objects from Gallagher et al.~(2006). The
  RQQ comparison sample is taken from Gibson et al.~(2008a) and is
  supplemented with luminous RQQs from Just et al.~(2007). Quasars
  from the DR5 Quasar Catalog of Schneider et al.~(2007) are shown as
  gray points. Our {\it Chandra} snapshot sample is biased toward
  luminous quasars as a consequence of the magnitude-limited selection
  method. The comparison samples have been constructed to overlap and
  bracket the BAL RLQs in optical luminosity and redshift.}
\end{figure}

\begin{figure}
\includegraphics[scale=0.75]{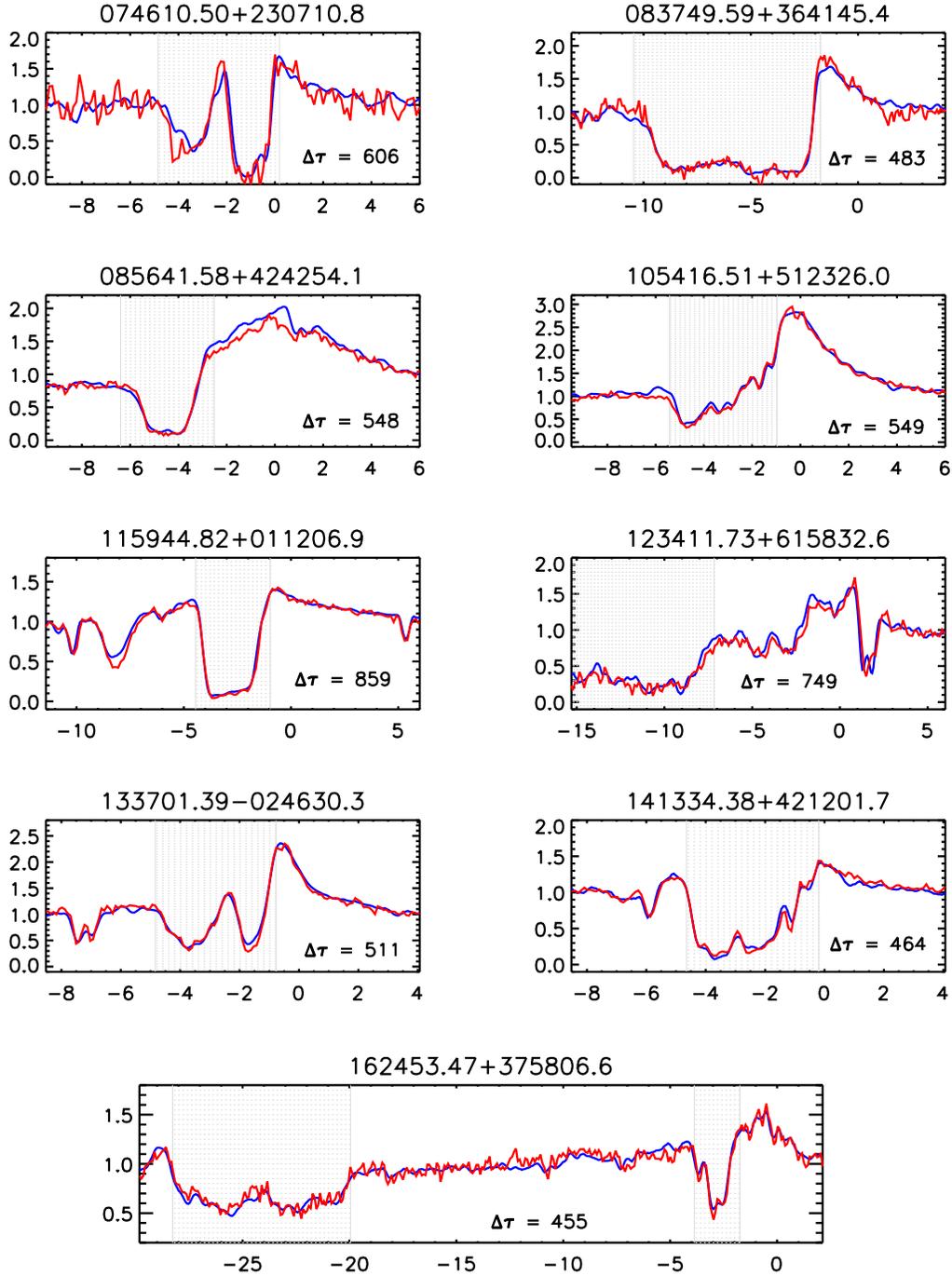} \figcaption{\small HET/LRS
  spectra (red lines) taken near the times of the {\it Chandra}
  snapshot observations, shown compared to the earlier epoch SDSS
  spectra (blue lines) matched to HET/LRS resolution. Each panel is
  labeled with the SDSS DR5 name as well as the rest-frame interval
  between observations (in days). The horizontal axis is velocity in
  1000~km~s$^{-1}$ and the vertical axis is normalized flux. The C~IV
  absorption regions are shaded gray. There is only minor BAL
  variability seen in these objects, indicating that variability does
  not significantly complicate a comparison of UV absorption
  properties to \hbox{X-ray} weakness.}
\end{figure}

\begin{figure}
\includegraphics[scale=0.4]{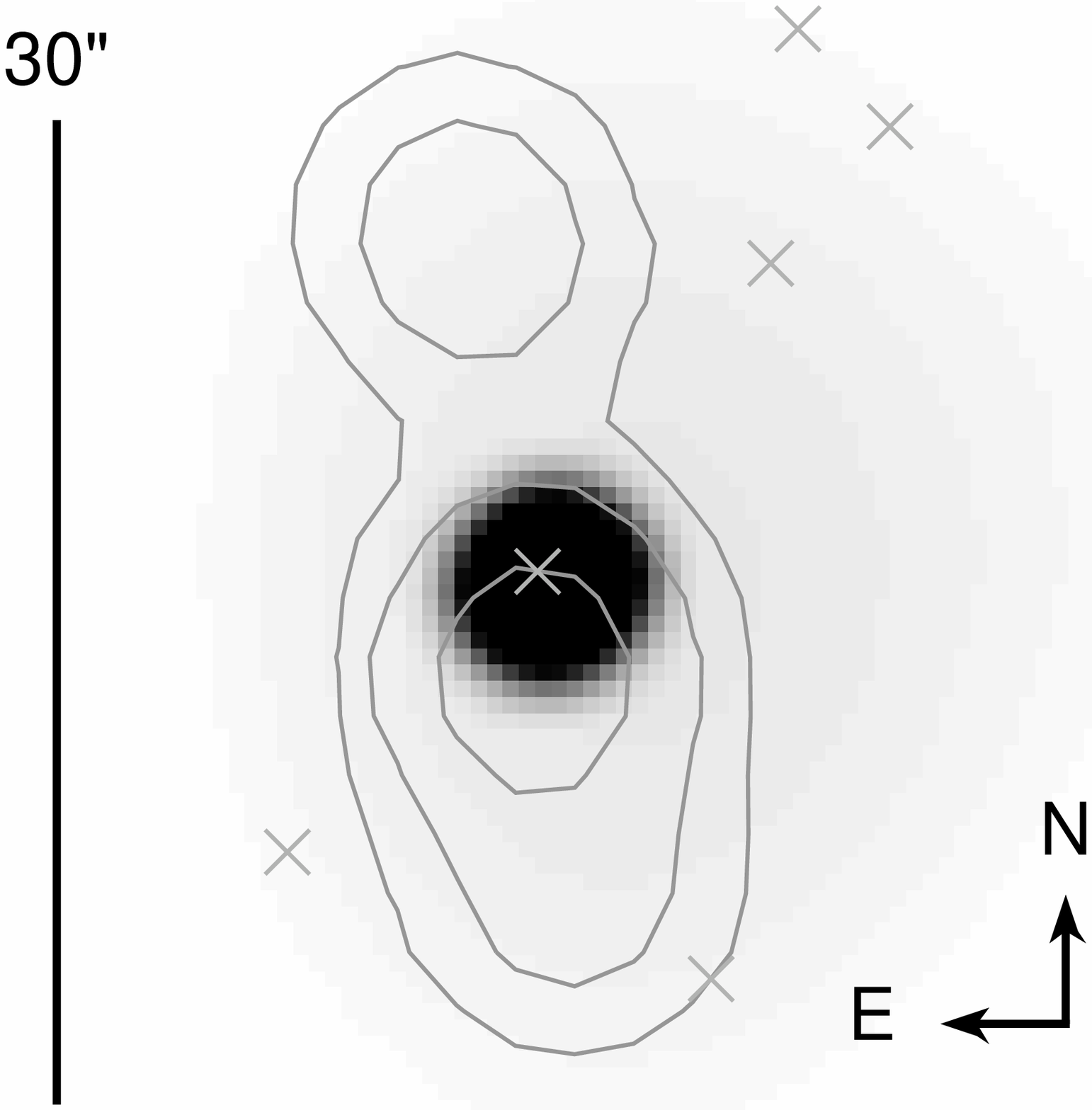}
\hspace{10.mm} \includegraphics[scale=0.4]{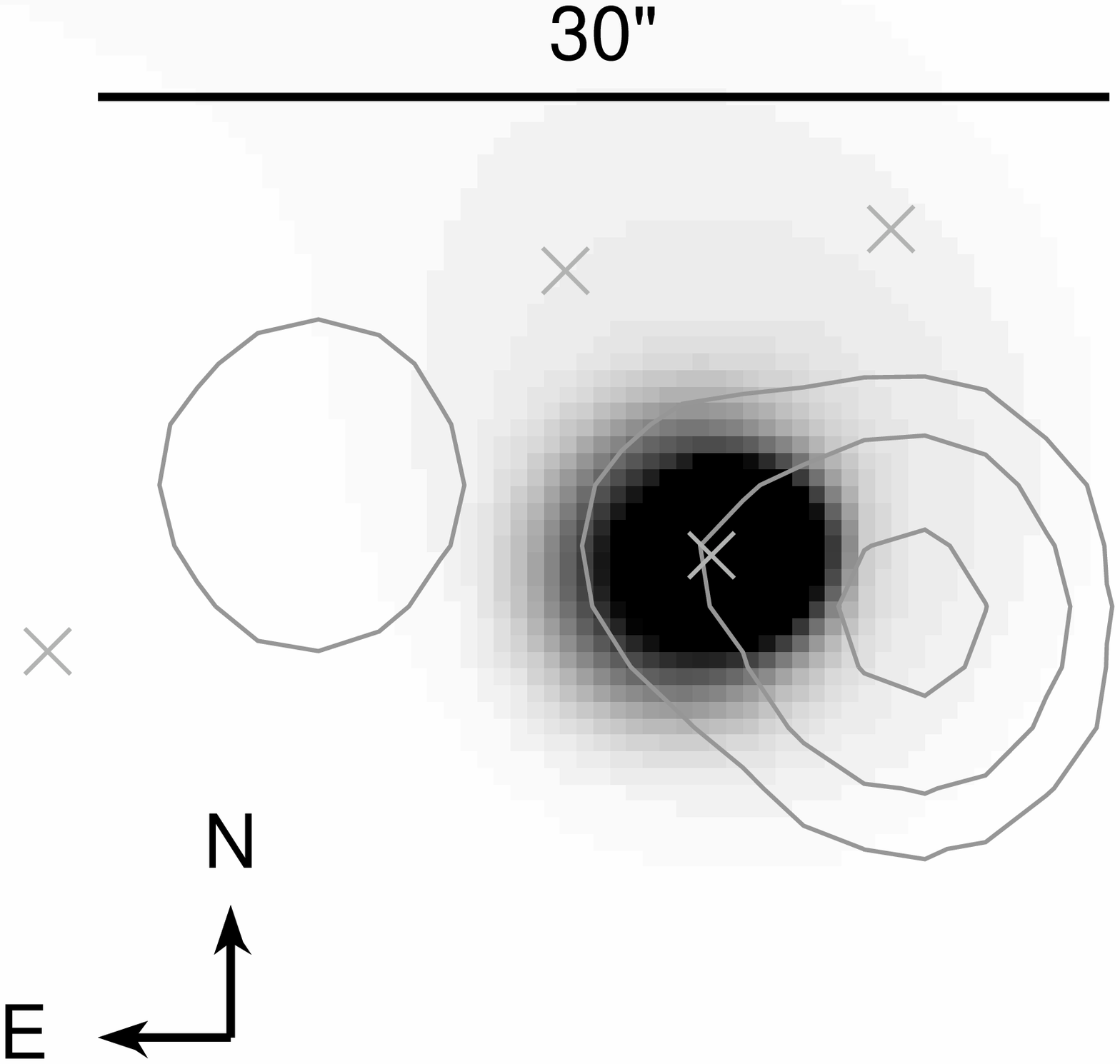}
\figcaption{\small {\it Chandra} images of the two BAL RLQs in the
  snapshot survey possessing extended radio emission. The left panel
  shows J102258.41+123429.7; the right panel shows
  J112506.95$-$001647.6. For both objects 30$''$ is
  $\simeq$250~kpc. Adaptively smoothed 0.5--8~keV images are plotted
  in grayscale with logarithmic scaling, overlaid with contours from
  the 5~GHz FIRST survey at levels of 2, 8, and
  32~mJy~beam$^{-1}$. Peak fluxes for the radio sources are $<85$\% of
  the integrated fluxes and the deconvolved major axes are
  $\sim3-5''$; FIRST apparently resolves these components. The crosses
  mark SDSS photometric sources within the field; none of these aligns
  with the apparent extended radio emission, further indicating that
  these are lobes rather than unrelated sources. }
\end{figure}

\begin{figure}
\includegraphics[scale=0.7]{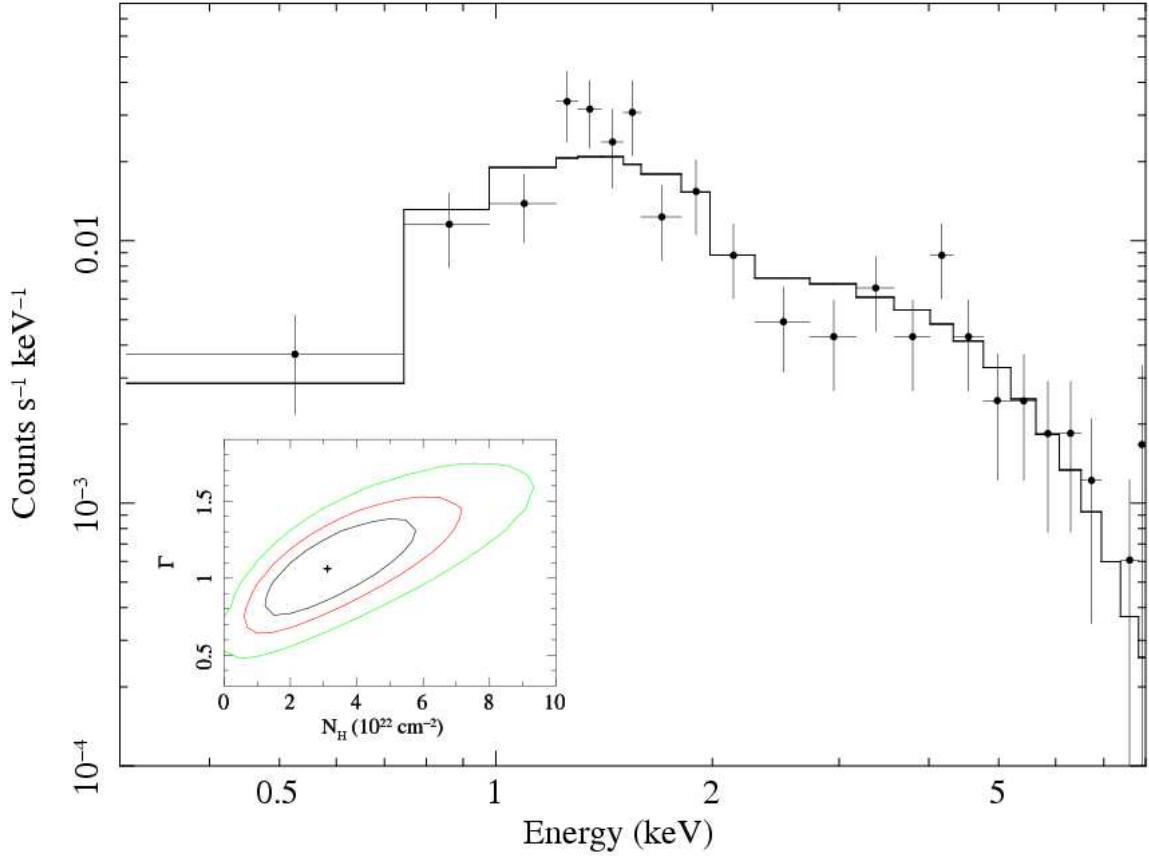}
\figcaption{\small {\it Chandra} spectrum of J115944.82+011206.9, the
X-ray brightest source in our snapshot sample with $\simeq$170 counts
from 0.3--8 keV ($\sim2\times$ that of the next-brightest snapshot BAL
RLQ). The plotted model has intrinsic absorption with column density
$N_{\rm H}=3.2^{+2.9}_{-2.1}\times10^{22}$~cm$^{-2}$ and a power-law
photon index of $\Gamma=1.06^{+0.35}_{-0.33}$. The fit was performed
using the $cstat$ statistic and the cosmetic binning is based on a
minimum significance of 3$\sigma$ within a maximum of 30 bins. The
inset shows the 68, 90, and 99\% confidence contours (for two
parameters of interest) for the photon index $\Gamma$ and intrinsic
column density $N_{\rm H}$.}
\end{figure}

\begin{figure}
\includegraphics[scale=0.70]{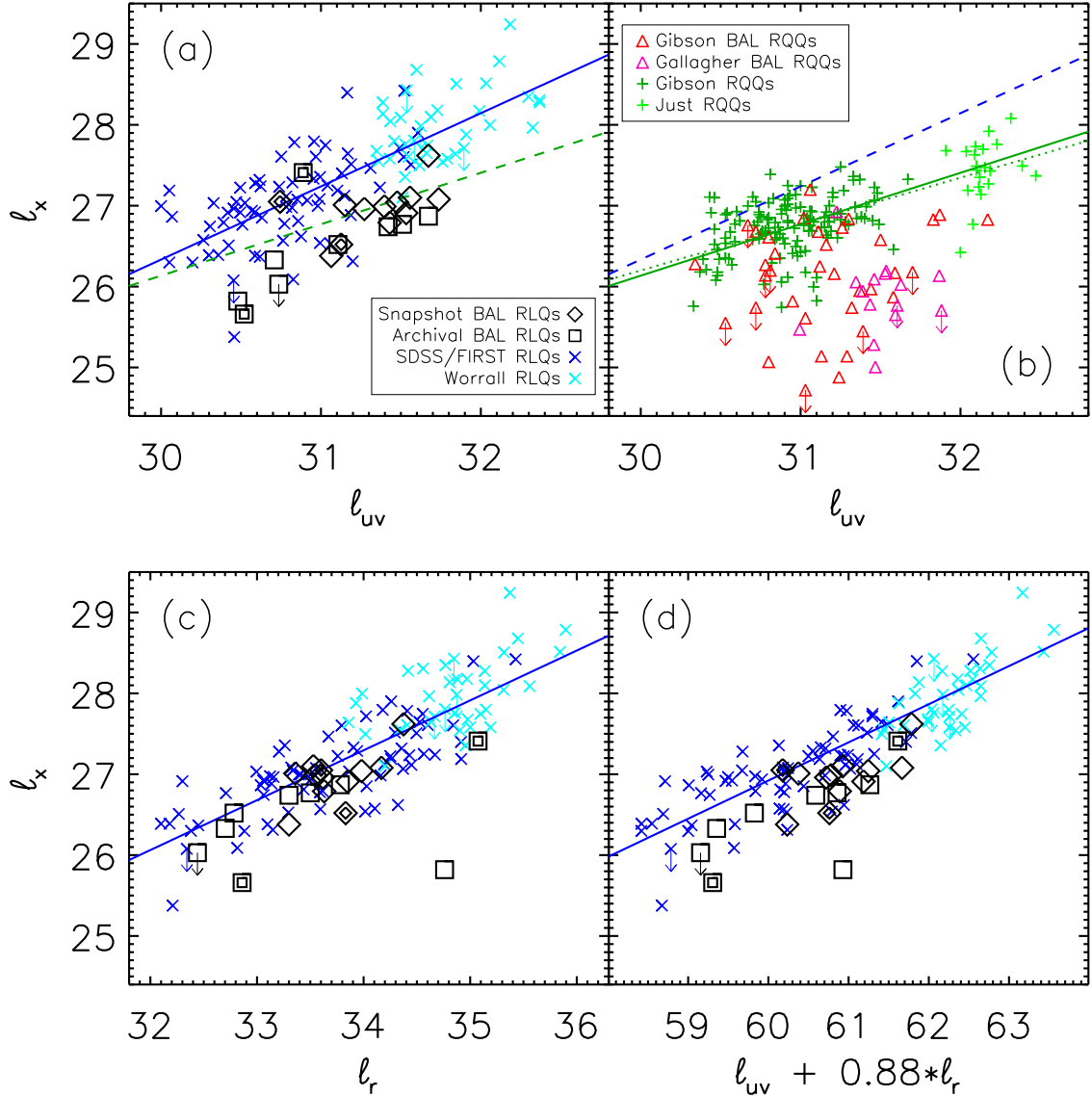} \figcaption{\small X-ray
  luminosities of BAL quasars compared to similar non-BAL
  quasars. Luminosities have units of log~ergs~s$^{-1}$~Hz$^{-1}$, at
  rest-frame frequencies of 5~GHz, 2500~\AA, and 2~keV for $l_{\rm
    r}$, $l_{\rm uv}$, and $l_{\rm x}$, respectively. Arrows indicate
  \hbox{X-ray} upper limits, and nested symbols are lobe-dominated BAL
  RLQs. Blue lines are best-fit correlations for non-BAL RLQs (taking
  $l_{\rm x}$ as the dependent variable) calculated using the Bayesian
  maximum-likelihood method of Kelly (2007). The solid green line
  shows the best-fit correlation for non-BAL RQQs that Just et
  al.~(2007) calculated with ASURV for a large sample of RQQs; fitting
  our comparison sample of RQQs yields a similar result (dotted green
  line). The $l_{\rm x}(l_{\rm uv})$ relation for RLQs/RQQs is also
  plotted as a dashed line in (b)/(a), illustrating the well-known
  tendency for RLQs to be \hbox{X-ray} brighter than comparable
  RQQs. BAL RLQs are \hbox{X-ray} weak relative to non-BAL RLQs with
  similar optical/UV luminosities (a) but not to the same degree as
  are BAL RQQs relative to non-BAL RQQs (b). BAL RLQs are also
  modestly \hbox{X-ray} weak relative to non-BAL RLQs with similar
  radio luminosities (c) and to non-BAL RLQs with both similar
  optical/UV and radio luminosities (d).}
\end{figure}

\begin{figure}
\includegraphics[scale=0.85]{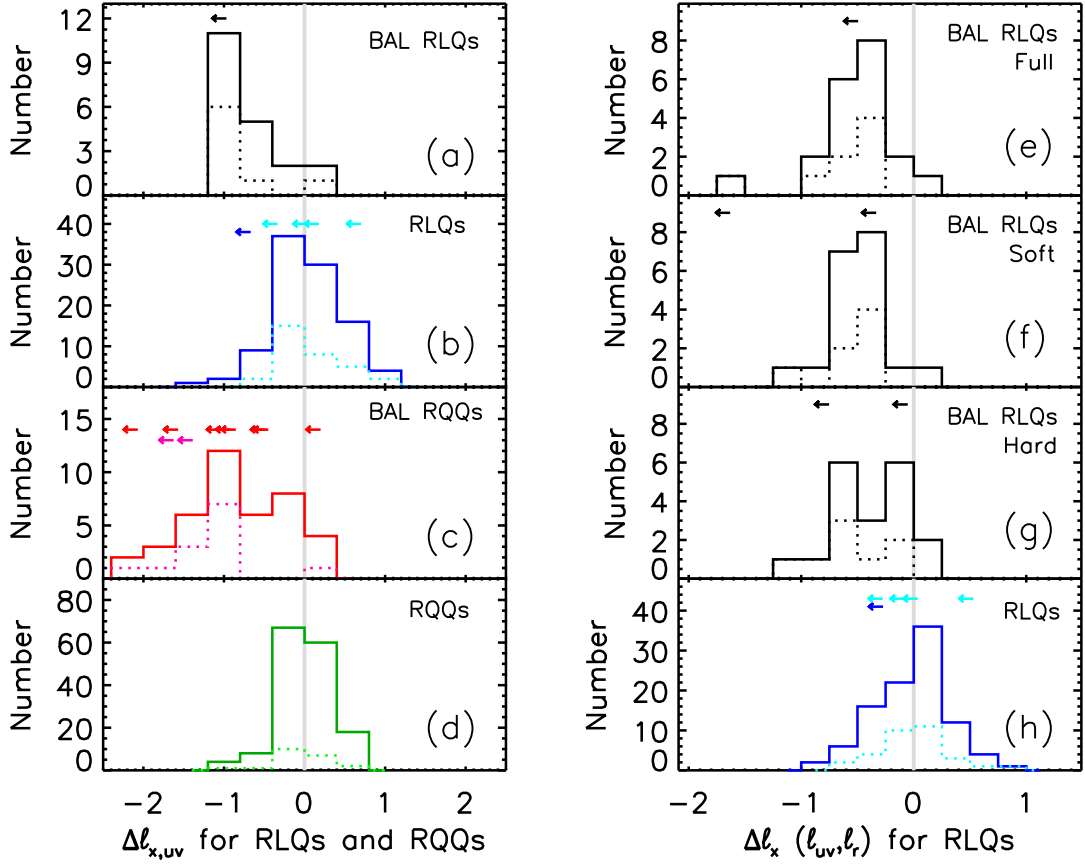} \figcaption{\small Histograms
  showing the distribution of the difference between actual and
  anticipated \hbox{X-ray} luminosity. Arrows indicate \hbox{X-ray} limits. Dotted
  histograms show subsamples: archival BAL RLQs (black), Worrall et
  al.~(1987) RLQs (cyan), Gallagher et al.~(2006) BAL RLQs (magenta),
  and Just et al.~(2007) RQQs (light green). The left column (a, b, c,
  d) shows ${\Delta}l_{\rm x,uv}$ calculated from optical/UV
  luminosities using the relations shown in Figures 9a and 9b. BAL
  RQQs reach more extreme values of \hbox{X-ray} weakness relative to non-BAL
  RQQs than do BAL RLQs relative to non-BAL RLQs with similar
  optical/UV luminosities. The right column (e, f, g, h) shows
  ${\Delta}l_{\rm x}$ calculated from both optical/UV {\it and\/}
  radio luminosities for RLQs, using the relation shown in Figure
  9d. \hbox{X-ray} luminosity is calculated using the full (0.5--8~keV), soft
  (0.5--2~keV), and hard (2--8~keV) counts. BAL RLQs are typically
  \hbox{X-ray} weaker than comparable non-BAL RLQs by a factor of
  2.0--4.5. The similarity of results derived using full, soft, and
  hard \hbox{X-ray} luminosities suggests simple absorption of the entire
  \hbox{X-ray} continuum source cannot provide a universal explanation for
  the \hbox{X-ray} weakness in BAL RLQs.}
\end{figure}

\begin{figure}
\includegraphics[scale=0.75]{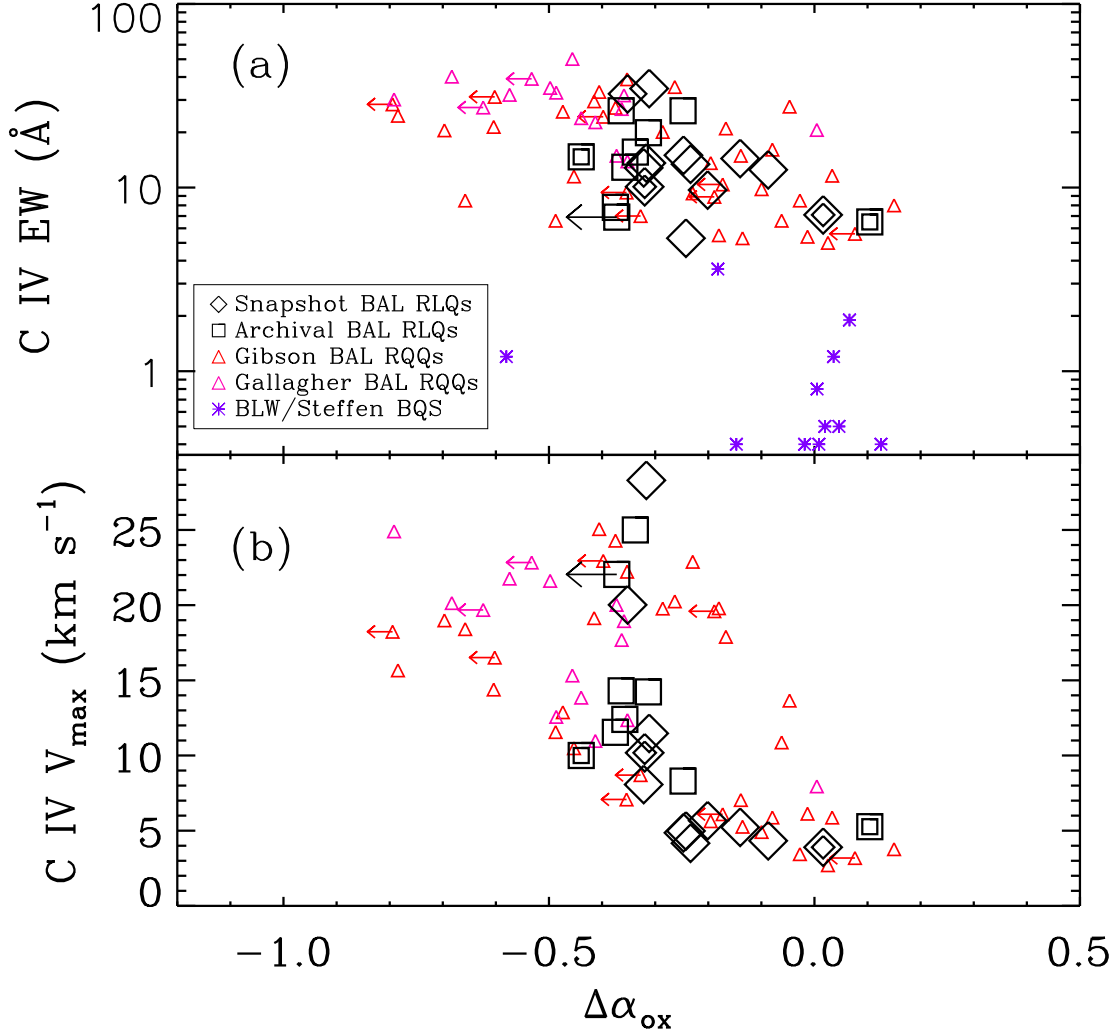} \figcaption{\small C~IV
  absorption properties as a function of relative \hbox{X-ray}
  luminosity, calculated using the relations shown in Figures 9a and
  9b and expressed in terms of ${\Delta}{\alpha}_{\rm ox} (l_{\rm
    uv})$, where \hbox{${\alpha}_{\rm ox} = 0.384\times(l_{\rm
      x}-l_{\rm uv})$}, for ease of comparison with previous
  work. Panel (a) shows C~IV EW and (b) shows maximum outflow
  velocity. Nested symbols are lobe-dominated BAL RLQs. The purple
  points in (a) are non-BAL RQQs from the BQS, with C~IV absorption
  values from Brandt, Laor, \& Wills (2000) and optical and
  \hbox{X-ray} luminosities from Steffen et al.~(2006). X-ray weakness
  appears more closely linked to absorption strength in BAL RQQs than
  in BAL RLQs; even BAL RLQs with extreme C~IV absorption properties
  do not have ${\Delta}{\alpha}_{\rm ox}<-0.5$, as do many strongly
  absorbed BAL RQQs.}
\end{figure}

\begin{figure}
\includegraphics[scale=0.8]{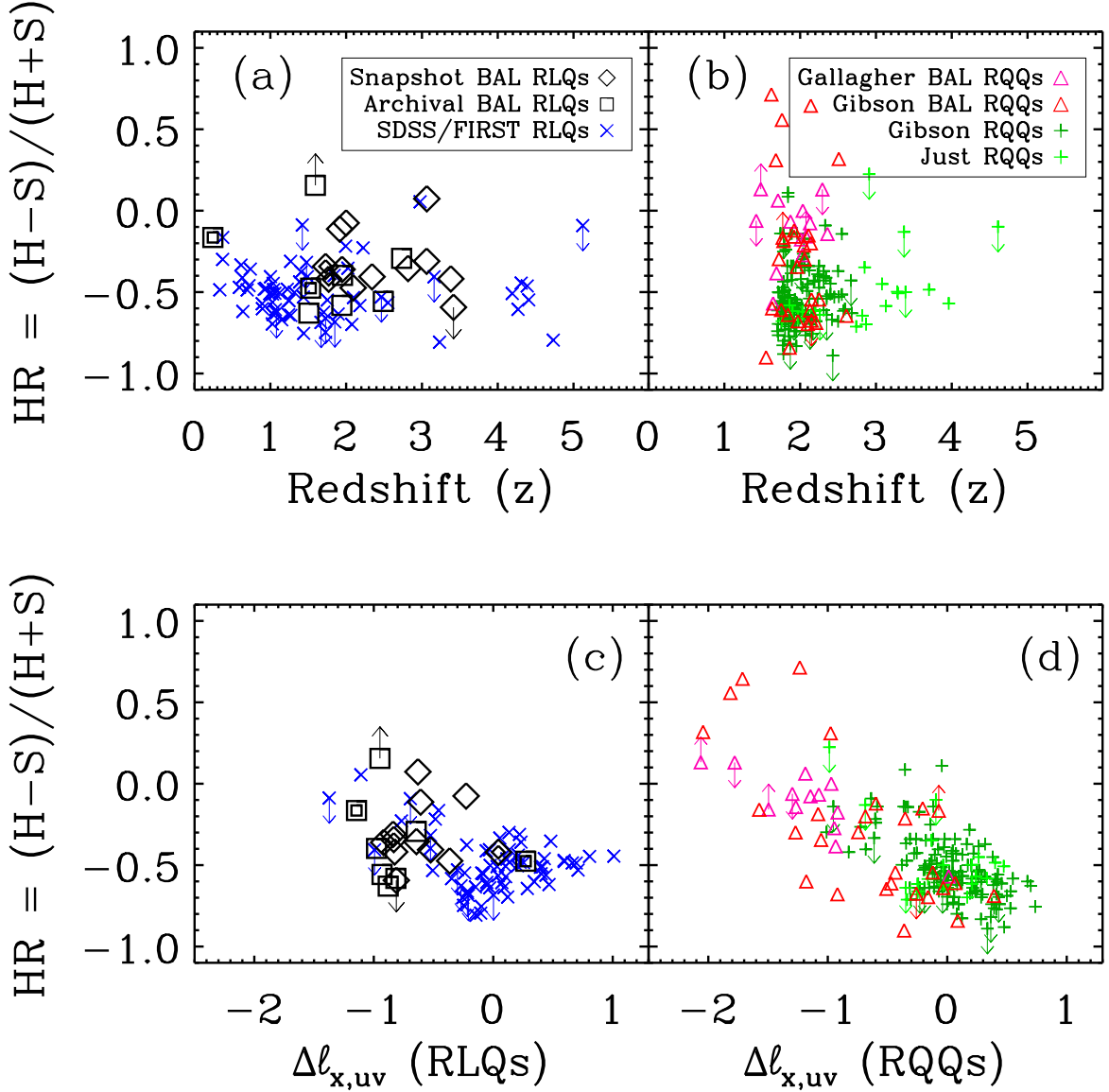} \figcaption{\small Hardness ratio
  plotted versus redshift for the sample of BAL RLQs (nested symbols
  are lobe-dominated BAL RLQs) and for comparison samples of RLQs (a)
  and for BAL RQQs and RQQs (b). The \hbox{X-ray} spectra of BAL RLQs
  are sometimes harder than those of typical non-BAL RLQs, but are
  often consistent. The \hbox{X-ray} spectra of BAL RQQs are often
  harder than for typical non-BAL RQQs, reaching extreme hardness
  ratios in some cases, and the distribution of hardness ratios for
  BAL RQQs is not statistically consistent with that of RQQs. Panels
  (c) and (d) show hardness ratio plotted versus ${\Delta}l_{\rm
    x,uv}$ calculated using the relations shown in Figures 9a and 9b;
  the \hbox{X-ray} weakness of BAL RQQs is linked to increasing
  intrinsic absorption of the continuum (illustrated via harder
  \hbox{X-ray} spectra), while many \hbox{X-ray} weak BAL RLQs do not
  have extremely hard \hbox{X-ray} spectra.}
\end{figure}

\begin{figure}
\includegraphics[scale=0.8]{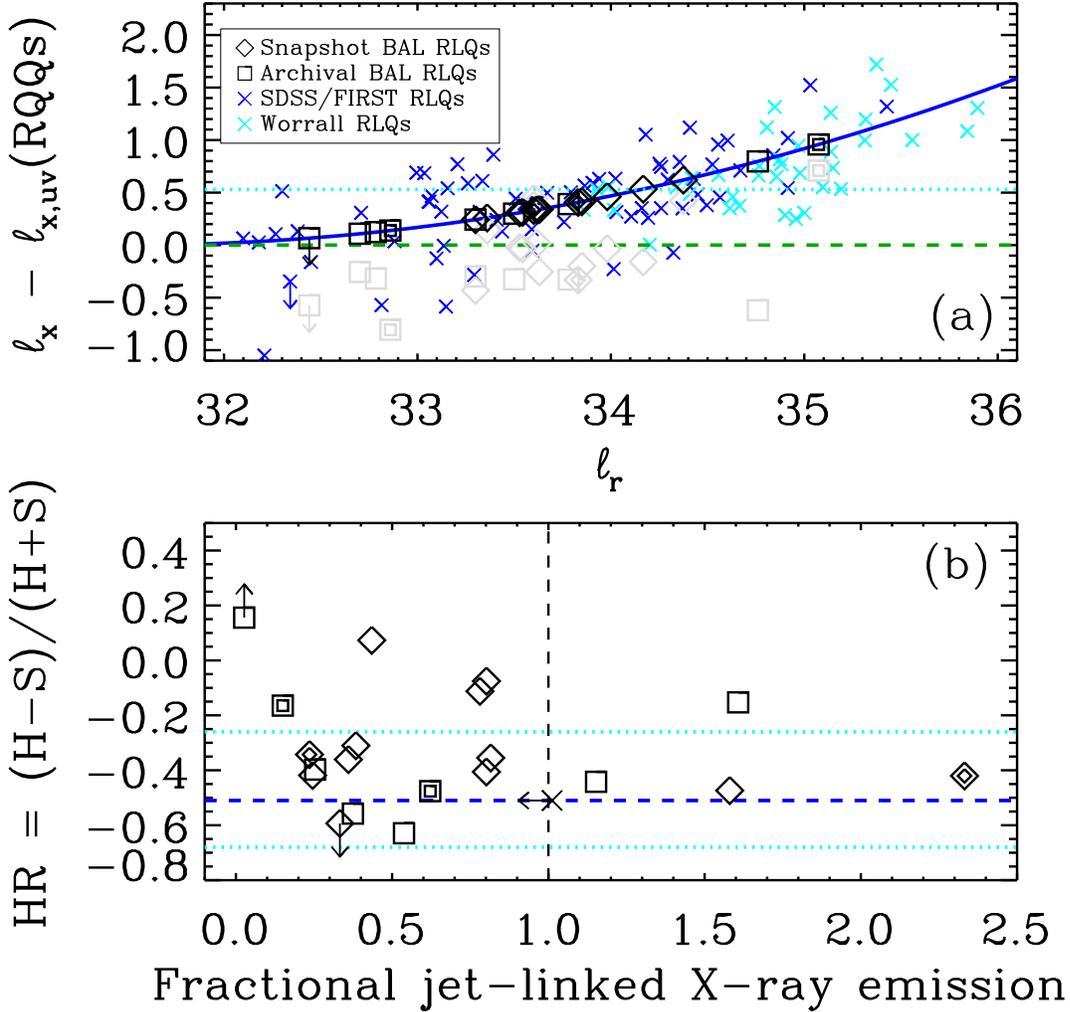} \figcaption{\small The top panel
  (a) shows the ratio of \hbox{X-ray} luminosity in RLQs to that of
  RQQs with comparable optical/UV luminosities, expressed in log
  units, as a function of radio luminosity. The median factor by which
  the comparison sample of RLQs are \hbox{X-ray} brighter than the
  comparison sample of RQQs is 3.4 (cyan dashed line); there is a
  trend (illustrated with the solid blue line) toward increasing X-ray
  brightness with increasing radio luminosity that likely reflects
  increasing jet dominance. BAL RLQs are plotted at their predicted
  (black) and observed (gray) \hbox{X-ray} luminosity ratios (nested
  symbols are lobe-dominated BAL RLQs). The bottom panel (b) shows the
  fraction of jet-linked \hbox{X-ray} emission in BAL RLQs relative to
  that expected for non-BAL RLQs with similar optical/UV and radio
  luminosities, assuming the disk/corona X-ray emission (predicted
  from the optical/UV luminosity using the RQQ relation) is reduced by
  a factor of 10, as is typical for BAL RQQs. Values for the
  fractional jet-linked emission near (or above) 1 would suggest the
  jet is unobscured, while values near 0.1 would indicate the jet is
  covered and reduced in intensity to a similar degree as is the
  disk/corona emission. The median, 10th, and 90th percentile values
  of hardness ratio for RLQs are also plotted (blue dashed and cyan
  dotted lines, respectively).}
\end{figure}

\begin{figure}
\includegraphics[scale=0.9]{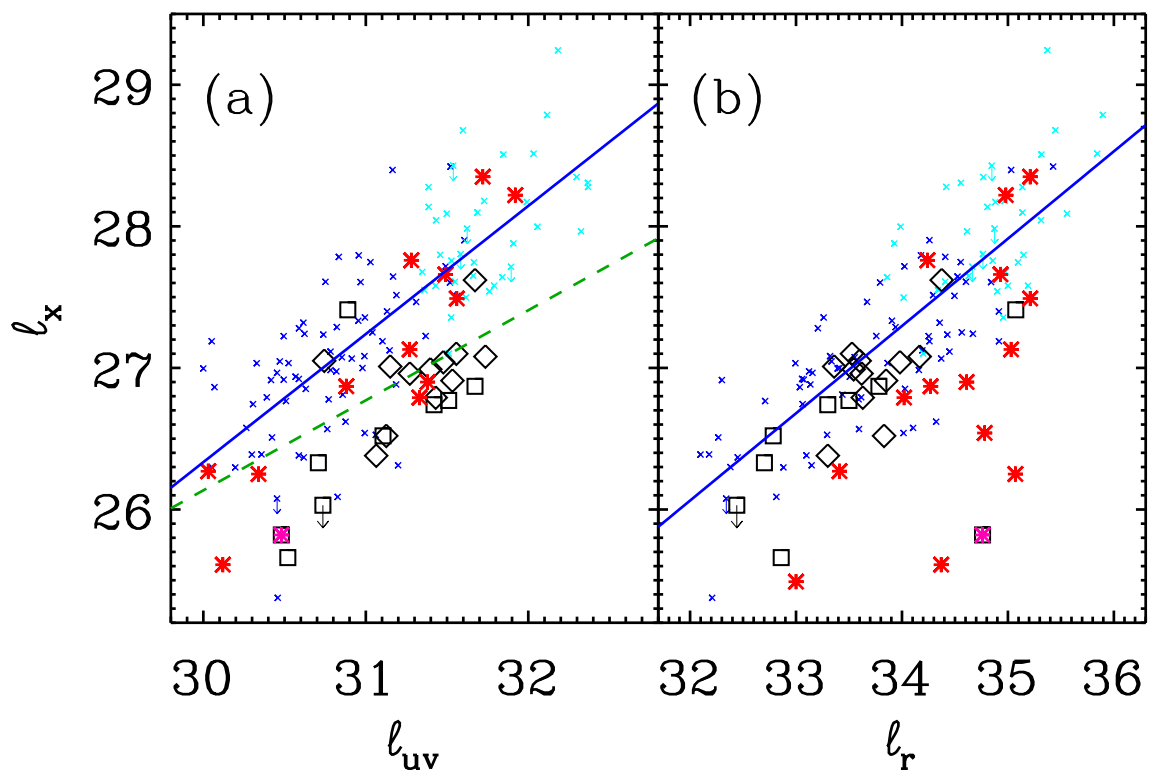} \figcaption{\small Comparison of
  the luminosities of GPS and CSS sources to those of BAL RLQs and
  non-BAL RLQs. Data for GPS and CSS sources (red stars) are from
  Siemiginowska et al.~(2008). The legend and caption for BAL RLQs and
  non-BAL RLQs is identical to Figures 9a and 9c for (a) and (b),
  respectively. The $l_{\rm x}(l_{\rm uv})$ relation for non-BAL RQQs
  from Figure 9b is shown as a dotted green line in (a). The CSS BAL
  RLQ J104834.24+345724.9 is indicated with a magenta star.}
\end{figure}

\end{document}